\documentclass{sty/IEEEtran}

\usepackage{amsmath}
\usepackage{amsfonts}
\usepackage{amssymb}
\usepackage{graphicx}
\ifCLASSOPTIONcompsoc
\usepackage[caption=false, font=normalsize, labelfont=sf, textfont=sf]{subfig}
\else
\usepackage[caption=false, font=footnotesize]{subfig}
\fi

\usepackage{cite}

\usepackage{threeparttable}
\usepackage{multirow}
\usepackage{times}
\usepackage{float}
\usepackage{enumerate}
\usepackage[none]{hyphenat}
\usepackage[normalem]{ulem}
\usepackage{color}
\usepackage{xcolor}
\usepackage{url}
\usepackage{tikz}
\usepackage{pgfplots}
\pgfplotsset{compat=newest}
\usetikzlibrary{positioning}

\usepackage{algorithm}
\usepackage{algorithmic}

\newcommand{\qed}{\nobreak \ifvmode \relax \else
    \ifdim\lastskip<1.5em \hskip-\lastskip
    \hskip22em plus0em minus0.5em \fi \nobreak
    \vrule height0.4em width0.3em depth0.25em\fi}

\newlength\fheight
\newlength\fwidth
\begin{document}

\title{An Adaptive Model Order Reduction Approach for the Finite Element Method in Time Domain in Electromagnetics}

\author{Ruth Medeiros and Valent\'{i}n de la Rubia
    \thanks{R. Medeiros and V. de la Rubia are with the Departamento de Matem\'{a}tica Aplicada a las {TIC}, ETSI de Telecomunicaci\'{o}n, Universidad Polit\'{e}cnica de Madrid, 28040 Madrid, Spain (e-mails: ruth.medeiros@upm.es; valentin.delarubia@upm.es).}
}

\maketitle

\begin{abstract}
    Time domain simulations are crucial for analyzing transient behavior and broadband responses in electromagnetic problems. However, conventional numerical methods such as finite element method in time domain (FEMTD) and finite difference time domain, can be computationally demanding due to their high-dimensional nature, making large-scale simulations impractical for design optimization and real-time analysis.

    This paper introduces TA-ROMTD, a time-adaptive reduced order model (ROM) for FEMTD simulations that significantly reduces computational costs while maintaining accuracy. The method alternates between FEMTD and a reduced order model in time domain (ROMTD), using an error estimator to detect when the ROMTD solution loses accuracy and switching back to FEMTD to update the ROM with new data. Thus, TA-ROMTD does not require prior knowledge of the problem, as the ROM is constructed on the fly using FEMTD data. A key feature of this approach is the use of a coarse time step during FEMTD time intervals, capturing essential system dynamics while minimizing computational overhead. By reducing the number of degrees of freedom, this method enables efficient electromagnetic simulations, making it a powerful tool for antenna and microwave circuit design.

    The efficiency of the TA-ROMTD strategy is demonstrated through numerical examples, including an antipodal Vivaldi antenna, a dielectric resonator antenna, a fully metallic dual-polarization frequency selective surface, and a mushroom-type electromagnetic bandgap structure. These cases show the capability of the proposed approach to achieve accurate solutions while significantly reducing the computation time from tens of hours to minutes.
\end{abstract}
\markboth{}{}

\begin{keywords}
    Computational electromagnetics (CEM), reduced order model (ROM), time-dependent Maxwell\textquotesingle s equations, microwave circuits and antennas, proper orthogonal decomposition (POD).
\end{keywords}
\IEEEpeerreviewmaketitle

\section{Introduction} \label{sec:introduction}
\PARstart{T}{he} increasing demand for fast and reliable design of antennas and microwave circuits has driven the development of advanced computational electromagnetic tools that aim to reduce simulation times without compromising the accuracy of the solution. Conventional numerical methods for solving electromagnetic problems in the time domain, such as finite difference time domain (FDTD) \cite{tan2007,liu2024,taflove2005,kenan2024,valverde2024}, discontinuous Galerkin in time domain (DGTD) \cite{angulo2015,yan2017,chen2018,chen2013,ren2024,dosopoulos2010MAG,dosopoulos2010TAP,dosopoulos2013,angulo2010,angulo2012}, and finite element method in time domain (FEMTD) \cite{lee1997,jiao2002,jiao2003,qing2012,bondeson2012,teixeira2008,wang2019,chen2020}, are widely used due to their ability to analyze complex interactions and structures. However, these methods often require substantial computational resources, especially for intricate geometries, large-scale problems, or fine computational meshes, making simulation time excessively long and complicating design optimization or real-time analysis.

Model order reduction (MOR) methodologies offer a powerful alternative by transforming high-dimensional electromagnetic problems into lower-dimensional representations without compromising accuracy. By reducing the number of degrees of freedom (DoFs) required to solve the problem \cite{benner2015}, MOR techniques significantly decrease the computational burden \cite{hess2013,delaRubia2009,delaRubia2018,feng2019,hochman2014,baltes2017,nicolini2019,codecasa2019,jiao2020,chellappa2021,ziegler2023,kappesser2024,szypulski2020,rewienski2016,fotyga2018} while preserving the essential dynamics of the system. This allows for faster simulations and more efficient exploration of diverse design possibilities.

Recent efforts have focused on developing reduced order models to efficiently solve Maxwell\textquotesingle s equations in the time domain \cite{li2018,li2018ieee,li2019,li2022,li2023,li2023nmpde,luo2016,nayak2024,nicolini2023,yu2024}. For instance, \cite{yu2024} proposed a data-driven approach for solving 2D time domain Maxwell\textquotesingle s equations for transverse magnetic waves. This method uses DGTD to generate high-dimensional solutions across multiple parameter sets, processes the data using high order dynamic mode decomposition (HODMD) \cite{leclainche2017} to create a reduced basis, and enables rapid simulations. However, this approach is computationally intensive due to the large number of simulations needed to create the database. Similar strategies are explored in \cite{li2022,li2023}.

In \cite{nayak2024}, a data-driven reduced order model (ROM) was introduced to accelerate FDTD simulations of cavity problems. This approach involves solving a subset of the simulation time with FDTD to collect data for the ROM construction, which is processed with dynamic mode decomposition (DMD) \cite{leclainche2017}. The solutions in the remaining time steps are extrapolated using the computed DMD-modes. This methodology has also been applied to speed up electromagnetic particle-in-cell plasma simulations \cite{nayak2024plasma}. On the other hand, in \cite{medeiros2024TAP}, the authors of this work proposed a reduced order model in time domain (ROMTD) to accelerate FEMTD simulations. The strategy uses larger FEMTD time steps during data collection to capture key dynamics efficiently, significantly reducing simulation time. However, for complex problems, even this approach can become computationally demanding because the FEMTD time interval must be sufficiently long to accurately capture the dynamics of the electromagnetic system. Another limitation of this method is its reliance on prior knowledge about the problem to determine an appropriate FEMTD time interval. If the selected interval is too short, the ROMTD approach will fail to accurately represent the solution.

To address these challenges, time-adaptive MOR techniques can be employed. These approaches alternate between the high-dimensional model and the ROM, significantly reducing computational costs by limiting the number of time steps solved by the high-dimensional model. For example, \cite{jane2023} demonstrated the effectiveness of a time-adaptive ROM for solving the electrochemical Doyle-Fuller-Newman model \cite{doyle1993} for lithium-ion battery cells, achieving a tenfold reduction in computational time for a 3D model. Furthermore, \cite{medeiros2024} introduced a time- and parameter-adaptive ROM strategy, yielding substantial cost reductions for battery cell design optimization.

This paper proposes a time-adaptive MOR strategy for FEMTD simulations, named TA-ROMTD, which can be extended to other numerical methods, such as FDTD and DGTD. The method begins by solving a short time interval using a relatively large time step to capture the key dynamics of the electromagnetic system and generate snapshots for ROM construction via proper orthogonal decomposition (POD) \cite{willcox2002}. The ROMTD then solves the FEMTD interval with a finer time step and continues the simulation until an error estimator indicates a loss of accuracy in the solution or until the final simulation time is reached. If the error estimator exceeds a predefined threshold, the model switches back to FEMTD for a brief period to collect additional data and update the ROM basis. Once updated, ROMTD continues solving the problem from the last ROMTD time step solved. It alternates between FEMTD and ROMTD as needed, whenever a loss of accuracy is detected. This adaptive approach reduces both the number of DoFs and the time steps solved using the time-consuming FEMTD, significantly improving computational efficiency compared to traditional methods. Additionally, the integration of an error estimator automates the resolution process, ensuring the ROM accurately captures the dynamics of the electromagnetic system without requiring prior knowledge of the problem.

The structure of this paper is as follows. Section~\ref{sec:problem_statement} provides an overview of the high-dimensional model and its numerical resolution. Section~\ref{sec:ROM} describes the proposed time-adaptive ROMTD strategy and details the formulation of the reduced order model. Section~\ref{sec:numericalResults} presents numerical results demonstrating the efficiency of the method in solving electromagnetic problems. Finally, Section~\ref{sec:conclusions} summarizes the key conclusions of this work.

\section{Problem Statement} \label{sec:problem_statement}

Maxwell\textquotesingle s equations describe the dynamics of electric and magnetic fields. In a linear, isotropic medium defined by a bounded, source-free analysis domain $\Omega \subset \mathbb{R}^3$, Maxwell\textquotesingle s equations in the time domain reduce to a system of two equations coupling the electric and magnetic fields. The boundary conditions define the behavior of these fields on surfaces characterized as perfect magnetic conductors (PMCs), perfect electric conductors (PECs), and excitation ports. Moreover, the system simplifies to a single governing equation for the electric field $\mathbf{E} (\mathbf{r}, t)$, expressed in V/m. Assuming the time interval $\mathcal{T} = (T_{i}, T_{f}]$, with $T_{i}$ and $T_{f}$ being the initial and final times, this equation is given by the following expression:
\begin{equation} \label{eq:strong_formulation}
    \begin{aligned}
        \nabla \times \bigg(\frac{1}{\mu} \nabla \times \mathbf{E}\bigg) + \sigma \frac{\partial \mathbf{E}}{\partial t} + \varepsilon \frac{\partial^{2} \mathbf{E}}{\partial t^{2}} = \mathbf{0}, & \quad (\mathbf{r}, t) \in \Omega \times \mathcal{T},
    \end{aligned}
\end{equation}
where $\mu$ denotes the magnetic permeability [H/m], $\sigma$ is the electrical conductivity [S/m], and $\varepsilon$ represents the dielectric permittivity [F/m]. The dielectric permittivity is expressed as $\varepsilon = \varepsilon_{0} \varepsilon_{r}$, where $\varepsilon_{0}$ and $\varepsilon_{r}$ represent the permittivity of free space [F/m] and the relative permittivity of the medium, respectively. Similarly, the magnetic permeability $\mu$ is given by $\mu = \mu_{0} \mu_{r}$, with $\mu_{0}$ being the permeability of free space [H/m], and $\mu_{r}$ the relative permeability of the medium. To simplify the expressions, the time and space dependencies of the electric field have been omitted. 

In addition, the boundary conditions considered are given by the following equations:
\begin{equation} \label{eq:bc}
    \begin{aligned}
        \mathbf{n} \times \mathbf{E} = \mathbf{0}, & \quad (\mathbf{r}, t) \in \Gamma_{\mathrm{PEC}} \times \mathcal{T}, \\
        \mathbf{n} \times \bigg(\frac{1}{\mu} \nabla \times \mathbf{E}\bigg) = \mathbf{0}, & \quad (\mathbf{r}, t) \in \Gamma_{\mathrm{PMC}} \times \mathcal{T}, \\
        \mathbf{n} \times \bigg(\frac{1}{\mu} \nabla \times \mathbf{E}\bigg) = - \frac{\partial \mathbf{J}}{\partial t}, & \quad (\mathbf{r}, t) \in \Gamma \times \mathcal{T}.
    \end{aligned}
\end{equation}
The boundary $\partial \Omega$ consists of PEC surfaces $\Gamma_{\mathrm{PEC}}$, PMC surfaces $\Gamma_{\mathrm{PMC}}$, and excitation ports $\Gamma$. Moreover, the excitation current $\mathbf{J} (\mathbf{r}, t)$ is applied at the ports and $\mathbf{n}$ represents the unit outward normal vector to the domain boundary.

In order to numerically solve the problem, equation \eqref{eq:strong_formulation} is reformulated in variational form, incorporating the boundary conditions \eqref{eq:bc}. Consequently, for each time $t \in \mathcal{T}$, the problem is expressed as follows:
\begin{equation} \label{eq:variational_formulation}
    \begin{aligned}
        \text{Find } & \mathbf{E} \in \mathcal{H} \text{ such that } \\
        & a(\mathbf{E}, \mathbf{v}) = l(\mathbf{v}), \ \forall \mathbf{v} \in \mathcal{H},
    \end{aligned}
\end{equation}
where $\mathcal{H}$ is a subspace of the Hilbert space $H(\mathrm{curl}, \Omega)$ \cite{kirsch2014}, consisting of functions $\mathbf{u} \in H(\mathrm{curl}, \Omega)$ for which $\mathbf{n} \times \mathbf{u} = \mathbf{0}$ on the PEC surfaces. The bilinear and linear forms $a$ and $l$ are defined as:
\begin{equation}
    \begin{aligned}
        &a(\mathbf{E}, \mathbf{v}) = \\
        &\int_{\Omega}\left( \frac{1}{ \mu } \nabla \times \mathbf{E} \cdot \nabla \times \mathbf{v} + \sigma \frac{\partial \mathbf{E}}{\partial t} \cdot \mathbf{v} + \varepsilon \frac{\partial^2 \mathbf{E}}{\partial t^2} \cdot \mathbf{v}\right) dV,
    \end{aligned}
\end{equation}
\begin{equation}
    l(\mathbf{v}) = - \int_{\Gamma} \frac{\partial \mathbf{J}}{\partial t} \cdot \mathbf{v} \ dS.
\end{equation}

The variational problem is solved using the finite element method (FEM) \cite{monk2003, jin2014}, as widely used in several studies \cite{lee1997,jiao2002,jiao2012}. The application of FEM to electromagnetic problems in the time domain is commonly known as FEMTD. This approach approximates the electric field $\mathbf{E}$ in a finite-dimensional subspace $\mathcal{H}_{h} \subset \mathcal{H}$. Therefore, considering a basis $\{\boldsymbol{w}_{j} (\mathbf{r})\}_{j=1}^{N_h}$ of the high-dimensional subspace $\mathcal{H}_{h}$ with dimension $N_{h}=\operatorname{dim}(\mathcal{H}_h)$, the approximation of the electric field $\mathbf{E}_{h} \in \mathcal{H}_{h}$ is expressed as a linear combination of these basis functions:
\begin{equation} \label{eq:FEM_decomposition}
    \mathbf{E}_{h}(\mathbf{r}, t) = \sum_{j=1}^{N_h} x_j(t) \ \boldsymbol{w}_{j} (\mathbf{r}),
\end{equation}
where $\mathbf{x} = (x_{1}, x_{2}, \ldots, x_{N_{h}})^{T}$ contains the unknown coefficients at each time $t \in \mathcal{T}$.

The time interval $\mathcal{T} = [T_{i}, T_{f}]$ is discretized into $N_{t}$ steps: $\{t^{0}, t^{1}, t^{2}, \ldots, t^{N_{t}}\}$. Moreover, the time value at each point of this partition is calculated using the relation $t^{n} = t^{n-1} + \Delta t$, where $\Delta t$ represents the time step size. The solution at time $t^{n}$, denoted as $\mathbf{x}^{n}$, is determined utilizing the Newmark-$\beta$ method for time integration \cite{jin2014,van2004}. The resulting discretized system is expressed as follows:
\begin{equation} \label{eq:FEM_time_discretization}
    \left( \frac{1}{4} \mathbf{S} + \frac{1}{\Delta t^{2}} \mathbf{T} + \frac{1}{2 \Delta t} \mathbf{U} \right) \mathbf{x}^{n} = \mathbf{f}^{n},
\end{equation}
where $\mathbf{f}^{n}$ is the time-dependent vector:
\begin{equation}
    \begin{aligned}
        \mathbf{f}^{n} = & \ \frac{1}{4} \mathbf{b}^{n} +\frac{1}{2} \mathbf{b}^{n-1} + \frac{1}{4} \mathbf{b}^{n-2} + \frac{2}{\Delta t^{2}} \mathbf{T} \mathbf{x}^{n-1} - \frac{1}{2} \mathbf{S} \mathbf{x}^{n-1} - \\
        & \frac{1}{\Delta t^{2}} \mathbf{T} \mathbf{x}^{n-2} + \frac{1}{2 \Delta t} \mathbf{U}\mathbf{x}^{n-2} - \frac{1}{4} \mathbf{S} \mathbf{x}^{n-2}.
    \end{aligned}
\end{equation}
The matrices $\mathbf{S}$, $\mathbf{T}$, and $\mathbf{U}$ are sparse square matrices of size $N_{h}$ and are obtained from spatial discretization. For $1 \leq i,j \leq N_{h}$, these matrices are defined as:
\begin{equation} \label{eq:FEM_matrices}
    \begin{aligned}
        S_{ij} = & \int_{\Omega} \frac{1}{\mu} \nabla \times \boldsymbol{w}_{j} \cdot \nabla \times \boldsymbol{w}_{i} \ dV, \\
        T_{ij} = & \int_{\Omega} \varepsilon \ \boldsymbol{w}_{j} \cdot \boldsymbol{w}_{i} \ dV, \\
        U_{ij} = & \int_{\Omega} \sigma \ \boldsymbol{w}_{j} \cdot \boldsymbol{w}_{i} \ dV,
    \end{aligned}
\end{equation}
and the right-hand side vector is:
\begin{equation} \label{eq:FEM_rhs}
    b_{i} = - \int_{\Gamma} \frac{\partial \mathbf{J}}{\partial t} \ \cdot \boldsymbol{w}_{i} \ dS.
\end{equation}
This approach is computationally demanding due to the high-dimensional system required to accurately capture the complex field variations across the discretized domain.

\section{Time-adaptive reduced order model} \label{sec:ROM}

\begin{figure*}[tbp]
    \centering
    \includegraphics[width=\textwidth]{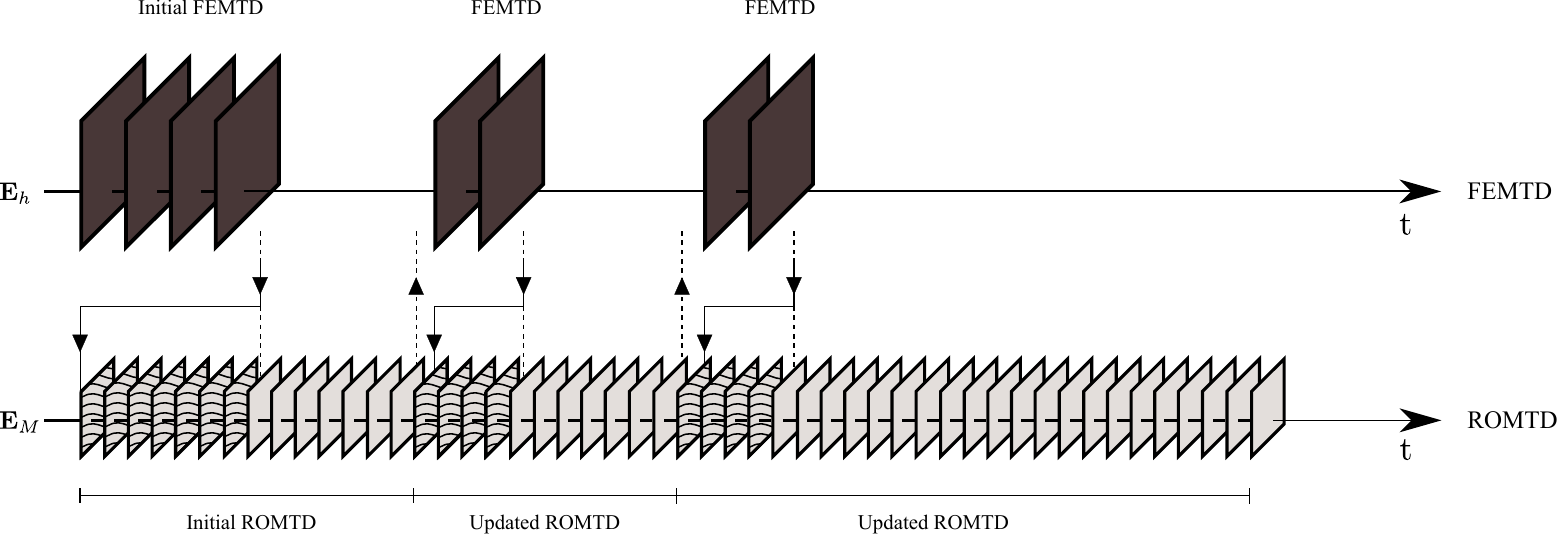}
    \caption{Schematic of the time-adaptive ROMTD strategy.}
    \label{fig:scheme_ROM}
\end{figure*}

\begin{figure*}[tbp]
    \centering
    \includegraphics[width=\textwidth]{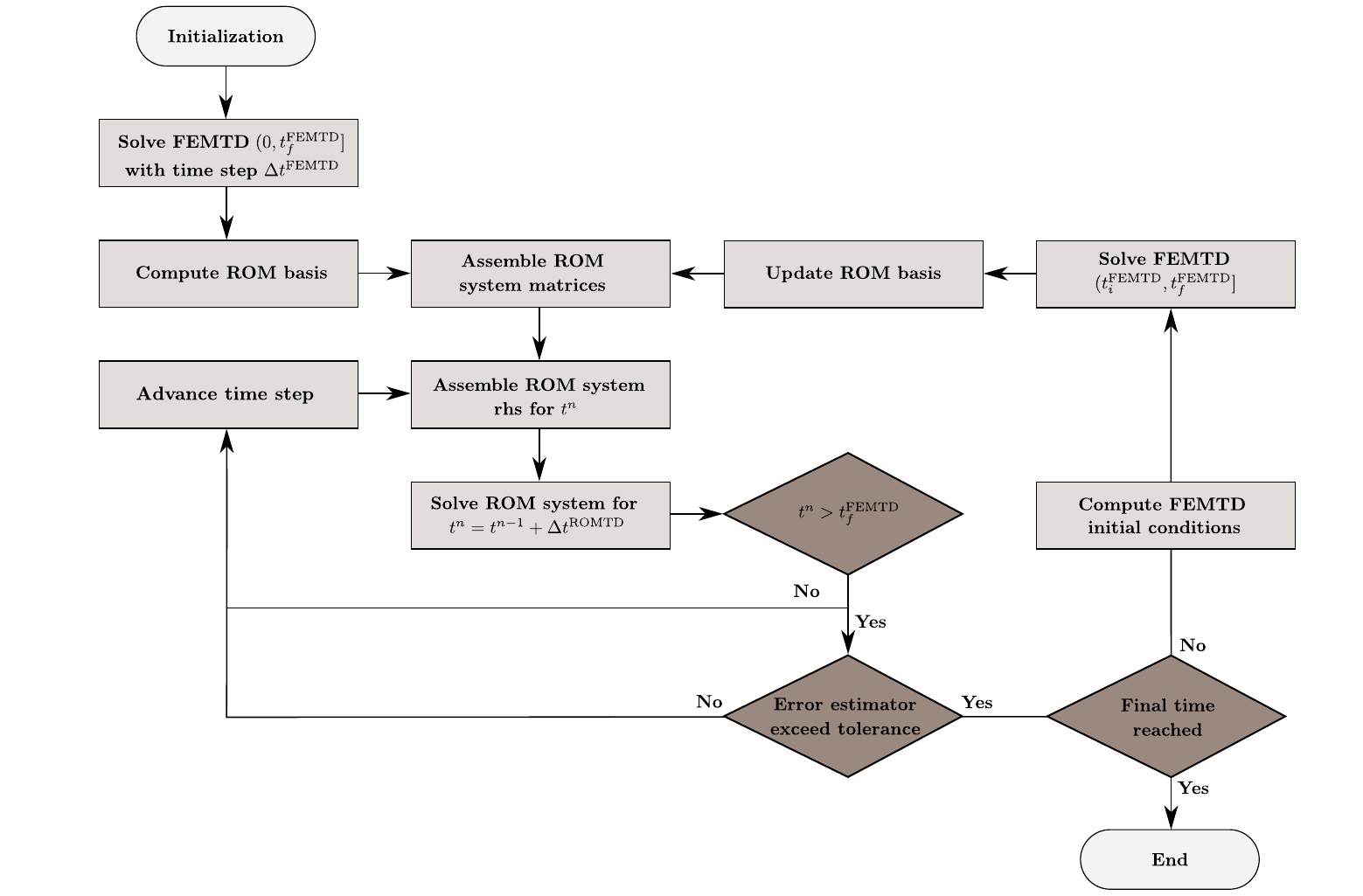}
    \caption{Flow diagram of the time-adaptive ROMTD strategy.}
    \label{fig:flowdiagram_taROMTD}
\end{figure*}

Reduced order models (ROMs) provide an efficient strategy to solve time domain Maxwell\textquotesingle s equations by reducing the computational cost of FEMTD simulations. By projecting the high-dimensional system onto a lower-dimensional space, ROMs enable faster simulations without losing accuracy. In a previous work \cite{medeiros2024TAP}, the authors proposed a ROM strategy called ROMTD for simulating the electromagnetic behavior of devices in the time domain. This approach requires to use FEMTD to solve a portion of the simulation time, generating a set of solutions, referred to as snapshots, to build the ROM. The time interval solved using FEMTD must be sufficiently long to capture the dynamics of the system. However, extending the interval length significantly increases the computational cost, which, for complex problems, can become prohibitively expensive. Additionally, the choice of the FEMTD time step is critical. As highlighted in \cite{medeiros2024TAP}, it can be larger than what is typically required for accurate problem resolution, further reducing the computational burden of the ROMTD approach. However, the time step should not be excessively large, as it must still ensure that the FEMTD solution accurately captures the dynamics of the system. This method requires some prior knowledge of the problem to select an FEMTD time interval that accurately captures the system dynamics. If the interval is too short, the reduced model may lack the necessary information to accurately represent the electromagnetic solution. Although the ROMTD strategy already accelerates the problem resolution without compromising accuracy, incorporating a time-adaptive strategy can enhance efficiency even further by reducing the number of FEMTD time step evaluations required.

A time-adaptive model order reduction (MOR) approach can significantly reduce the computational cost of solving problem \eqref{eq:variational_formulation} using the FEMTD method. This strategy, referred to as time-adaptive ROMTD (TA-ROMTD) and illustrated in Fig.~\ref{fig:scheme_ROM}, alternates between FEMTD and ROMTD solvers. Initially, the FEMTD approach solves the problem over a brief time interval using a large time step, $\Delta t^{\mathrm{FEMTD}}$, chosen fine enough to capture the dynamics of the system but large enough to avoid excessive computational cost. The solution of this initial FEMTD simulation is used to calculate the ROM basis via \emph{proper orthogonal decomposition} (POD). Then, the ROMTD approach is employed to solve from the first time step solved by FEMTD onward, using a finer time step $\Delta t^{\mathrm{ROMTD}}$. During subsequent time integration, the ROMTD is the primary solver, with periodic updates of the ROM basis using the FEMTD whenever the ROM loses accuracy. For each ROMTD update, FEMTD solves a small time interval using the large time step, $\Delta t^{\mathrm{FEMTD}}$. Notably, ROMTD always uses the finer time step, $\Delta t^{\mathrm{ROMTD}}$, which is ideally much smaller than $\Delta t^{\mathrm{FEMTD}}$. These FEMTD evaluations supply essential data to update and maintain the ROM precision and accuracy.

The remaining of this section is structured as follows: Section \ref{sec:ROM_formulation} presents the formulation of the ROM problem, while Section \ref{sec:ROM_adaptiveStrategy} describes the adaptation strategy, including the error estimator and the technique used to update the ROM basis during time integration.

\subsection{ROMTD formulation} \label{sec:ROM_formulation}
The ROMTD methodology approximates the FEMTD solution $\mathbf{E}_{h} \in \mathcal{H}_{h}$ using a function $\mathbf{E}_{M}$ from a lower-dimensional subspace $\mathcal{H}_{M}$ of dimension $\operatorname{dim}(\mathcal{H}_{M}) = M \ll \operatorname{dim}(\mathcal{H}_{h})$. As mentioned above, the basis determining the subspace $\mathcal{H}_{M}$ is constructed using POD \cite{sirovich1987}. Consider the electric field solutions at $m$ time steps obtained by means of the FEMTD approach, namely,
\begin{equation}
    \mathbf{E}^{1}, \mathbf{E}^{2}, \ldots, \mathbf{E}^{m}.
\end{equation}
All snapshots are arranged into a matrix to compute the POD utilizing the well-known \emph{snapshot method}, as explained in Appendix \ref{sec:POD}. The first $M$ \emph{POD-modes} are used as a basis for the reduced space $\mathcal{H}_{M}$, with $M < m$.

The approximated ROMTD solution is expressed as a linear combination of the POD-modes:
\begin{equation} \label{eq:ROM}
    \mathbf{E}_{h}(\mathbf{r}, t) \simeq \mathbf{E}_{M}(\mathbf{r}, t) = \sum_{j=1}^{M} \tilde{x}_{j}(t) \ \boldsymbol{\psi}_{j} (\mathbf{r}),
\end{equation}
where the dimension $M$ must be much smaller than $N_{h}$ to ensure computational efficiency.

Substituting this representation into the variational formulation \eqref{eq:variational_formulation} and using $\boldsymbol{\psi}_j$ as test functions yields a discretized system of ordinary differential equations. Time integration follows the Newmark-$\beta$ method, analogously to the FEMTD approach previously described. Assuming that the solution $\mathbf{\tilde{x}} (t) = (\tilde{x}_{1} (t), \tilde{x}_{2} (t), \ldots, \tilde{x}_{M} (t))^{T}$ at time steps $t^{n-1}$ and $t^{n-2}$ is known, denoted as $\mathbf{\tilde{x}}^{n-1}$ and $\mathbf{\tilde{x}}^{n-2}$, the solution at $t^{n}$ is obtained by solving the following system:
\begin{equation} \label{eq:ROM_time_discretization}
    \left( \frac{1}{4} \mathbf{\tilde{S}} + \frac{1}{\Delta t^{2}} \mathbf{\tilde{T}} + \frac{1}{2 \Delta t} \mathbf{\tilde{U}} \right) \mathbf{\tilde{x}}^{n} = \mathbf{\tilde{f}}^{n},
\end{equation}
where the right-hand side vector is given by:
\begin{equation}
    \begin{aligned}
        \mathbf{\tilde{f}}^{n} = & \ \frac{1}{4} \mathbf{\tilde{b}}^{n} +\frac{1}{2} \mathbf{\tilde{b}}^{n-1} + \frac{1}{4} \mathbf{\tilde{b}}^{n-2} + \frac{2}{\Delta t^{2}} \mathbf{\tilde{T}} \mathbf{\tilde{x}}^{n-1} - \frac{1}{2} \mathbf{\tilde{S}} \mathbf{\tilde{x}}^{n-1} \\
        & - \frac{1}{\Delta t^{2}} \mathbf{\tilde{T}} \mathbf{\tilde{x}}^{n-2} + \frac{1}{2 \Delta t} \mathbf{\tilde{U}}\mathbf{\tilde{x}}^{n-2} - \frac{1}{4} \mathbf{\tilde{S}} \mathbf{\tilde{x}}^{n-2}.
    \end{aligned}
\end{equation}
The matrices involved in this system are dense, square, and have dimension $M \ll N_{h}$. Let $\boldsymbol{\Psi}$ be the matrix whose columns are the POD-modes at the degrees of freedom of the FEMTD problem. With this representation, the matrices and vector of the reduced system \eqref{eq:ROM_time_discretization} can be computed as follows:
\begin{equation}
    \begin{aligned}
        \mathbf{\tilde{S}} = & \boldsymbol{\Psi}^{T} \mathbf{S} \ \boldsymbol{\Psi}, \\
        \mathbf{\tilde{T}} = & \boldsymbol{\Psi}^{T} \mathbf{T} \ \boldsymbol{\Psi}, \\
        \mathbf{\tilde{U}} = & \boldsymbol{\Psi}^{T} \mathbf{U} \ \boldsymbol{\Psi}, \\
        \mathbf{\tilde{b}} = & \boldsymbol{\Psi}^{T} \mathbf{b},
    \end{aligned}
\end{equation}
where $\mathbf{S}$, $\mathbf{T}$ and $\mathbf{U}$ correspond to the high-dimensional system matrices defined in expressions \eqref{eq:FEM_matrices}, while $\mathbf{b}$ represents the right-hand side of the high-dimensional model in expression \eqref{eq:FEM_rhs}. For a detailed derivation of the ROMTD formulation, refer to \cite{medeiros2024TAP}.

\subsection{Time-adaptive strategy} \label{sec:ROM_adaptiveStrategy}
Ensuring the precision of the solution obtained using the reduced order model is crucial for reliable results. Error estimators are an effective technique to achieve this, as they quantify the deviation of the ROMTD solution from the high-dimensional model solution. Thus, they provide a mechanism for determining when it is necessary to switch from the ROMTD back to the FEMTD to obtain new data and update the ROM basis.

The accuracy of the time-adaptive ROMTD approach is evaluated using the following error estimator:
\begin{equation} \label{eq:error_estimator}
    E_{\mathrm{err}} = \sqrt{  \sum_{n = N}^{M} (\tilde{x}_{n})^{2} \bigg/ \sum_{n = 1}^{M} (\tilde{x}_{n})^{2}  }, 
\end{equation}
where $N < M$ identifies the less relevant POD-modes, whose energy contribution is less than a thousandth of the total energy of all POD-modes. The estimator \eqref{eq:error_estimator} monitors the high-order POD coefficients in $\mathbf{\tilde{x}}$ during ROMTD time integration. When the ROM subspace can no longer adequately represent the FEMTD solution, the error estimator value increases, signaling the need for higher-order POD-modes to address errors in the electric field representation. This increase indicates that high-order modes have become relevant for accurately representing the field, thus necessitating an update of the ROM basis. This error estimation has previously been utilized in \cite{terragni2011, rapun2015}.

As introduced earlier, each ROM begins by solving the same time interval previously computed by the FEMTD solver, but with a finer time step $\Delta t^{\mathrm{ROMTD}}$. After solving this interval, the error estimator is computed at each time step to evaluate the accuracy of the ROMTD solution. When the estimated error exceeds a predefined threshold, the model switches back to FEMTD for a few time steps before returning to ROMTD, collecting new data to be used to uplift the ROM basis. The updated ROMTD continues to solve the problem until the final simulation time is reached or until the error estimator threshold is exceeded again. If the error estimator remains below the specified threshold throughout the simulation, FEMTD is only required for a brief initial time interval. A flow diagram illustrating this time-adaptive ROMTD strategy is detailed in Fig. \ref{fig:flowdiagram_taROMTD}.

To update the ROM basis, a technique is employed that combines newly obtained FEMTD data with the existing ROM basis. This process involves solving a short time interval using FEMTD and then performing POD on a matrix composed of POD-modes from both the previous and newly obtained FEMTD snapshots. Therefore, this technique generates an orthonormal basis, removing redundancy and retaining only the essential information. The procedure for updating the reduced basis is explained in more detail below.

The existing ROM basis consists of $M$ POD-modes:
\begin{equation}
    \mathcal{B} = \{ \boldsymbol{\psi}_{1} (\mathbf{r}), \boldsymbol{\psi}_{2} (\mathbf{r}), \ldots, \boldsymbol{\psi}_{M} (\mathbf{r}) \},
\end{equation}
while the latest FEMTD snapshots are used to generate $\hat{M}$ new POD-modes, denoted as
\begin{equation}
    \hat{\mathcal{B}} = \{ \hat{\boldsymbol{\psi}}_{1} (\mathbf{r}), \hat{\boldsymbol{\psi}}_{2} (\mathbf{r}), \ldots, \hat{\boldsymbol{\psi}}_{\hat{M}} (\mathbf{r}) \}.
\end{equation}

\begin{algorithm}[tbp!]
    \caption{Algorithm for updating the ROM Basis}
    \begin{enumerate}
        \item Compute POD-modes from the latest FEMTD time interval:
            \begin{equation*}
                \hat{\boldsymbol{\psi}}_{1} (\mathbf{r}), \hat{\boldsymbol{\psi}}_{2} (\mathbf{r}), \ldots, \hat{\boldsymbol{\psi}}_{\hat{M}} (\mathbf{r})
            \end{equation*}
        \item Construct a matrix by combining the new POD-modes with the previous ROM basis:
        \begin{equation*}
            \begin{bmatrix}
                \vert &  & \vert & \vert & & \vert\\
                \boldsymbol{\psi}_{1} & \ldots & \boldsymbol{\psi}_{M} & \hat{\boldsymbol{\psi}}_{1} & \ldots & \hat{\boldsymbol{\psi}}_{\hat{M}}\\
                \vert & & \vert & \vert & & \vert
            \end{bmatrix}
        \end{equation*}
        \item Carry out POD on this matrix to obtain an orthonormal basis.
        \item Truncate the obtained POD-modes to form the updated ROM basis.
    \end{enumerate}
    \label{alg:adaptation}
\end{algorithm}

The POD-modes from $\mathcal{B}$ and $\hat{\mathcal{B}}$ are combined into a single matrix, which is then processed using POD. As a result, an updated ROM basis is obtained. This update process removes redundant information between the two bases and incorporates the relevant information from the new FEMTD data. The full procedure for updating the ROM basis is outlined in Algorithm~\ref{alg:adaptation}.

Additionally, when the ROM basis is updated with new FEMTD data, the solutions from the last two time steps of the previous ROMTD must be recalculated for the updated ROM basis, due to the Newmark-$\beta$ method chosen for time integration. This process begins by converting these solutions back into the high-dimensional subspace using the previous set of POD-modes and expression \eqref{eq:ROM}. After the ROM basis has been updated with the new FEMTD data, the recalculated high-dimensional solutions are projected onto the updated reduced subspace, providing the necessary initial conditions for the new ROMTD.

The proposed time-adaptive ROMTD strategy achieves both computational efficiency and accuracy by updating the ROM basis whenever the ROMTD solution significantly deviates from the FEMTD solution. To automate the simulation using this methodology, three control parameters are required: the fraction of the total simulated time that the initial FEMTD solves, the POD-modes truncation tolerance, and the error estimator threshold. The proposed approach uses error estimation to ensure the ROM accurately represents the FEMTD solution. When the ROM loses precision, incorporating the new FEMTD data into the ROM basis through the described adaptation strategy removes redundant information and enriches the ROM basis with essential new data.

\section{Numerical Results} \label{sec:numericalResults}

This section shows the capabilities of the time-adaptive ROMTD strategy described above to efficiently solve problem \eqref{eq:variational_formulation}, thereby avoiding the high computational costs of using the FEMTD methodology. In particular, four different examples are analyzed: an antipodal Vivaldi antenna, a dielectric resonator antenna, a fully metallic dual-polarization frequency selective surface, and a mushroom-type electromagnetic bandgap structure.

For each example, the time-adaptive ROMTD results are compared with those obtained using FEMTD to show the accuracy of the proposed approach. Furthermore, the time domain solutions are post-processed using the fast Fourier transform (FFT) to convert them to the frequency domain for further analysis \cite{grivet2015}. This conversion enables direct comparison with solutions computed by solving problem \eqref{eq:strong_formulation} in the frequency domain, ensuring that the time domain approach accurately captures relevant frequency domain behavior.

The numerical results were obtained utilizing an in-house C++ code specifically developed to use the FEM for solving Maxwell\textquotesingle s equations. The implementation employed second-order first-family N\'ed\'elec's finite elements \cite{nedelec1980, ingelstrom2006} on second-order tetrahedral meshes created with \texttt{Gmsh} \cite{geuzaine2009}. All simulations were carried out on a workstation equipped with two 3.00-GHz Intel Xeon E5-2687W v4 processors and 512 GB of RAM.

\subsection{Antipodal Vivalvi Antenna} \label{sec:results_antipodalVivaldi}

\begin{figure}[tbp]
    \centering
    \includegraphics[width=0.9\linewidth]{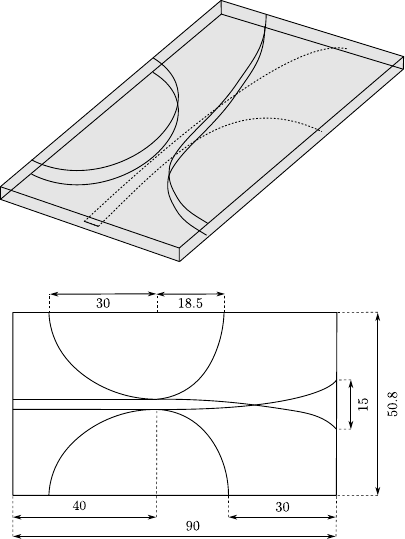}
    \caption{Representation of the geometry of the antipodal Vivaldi antenna, with dimensions specified in millimeters.}
    \label{fig:AntipodalVivaldiAntenna_geometry}
\end{figure}

This section presents the application of the time-adaptive ROMTD strategy to analyze the electromagnetic behavior of an antipodal Vivaldi antenna fed by a stripline, as proposed in \cite{lou2005}. Fig. \ref{fig:AntipodalVivaldiAntenna_geometry} illustrates a schematic of the antenna. The design has a $3$ mm wide stripline in the center of a $3.15$ mm thick dielectric plate, which gradually expands to create a radiating fin. Moreover, ground planes on the upper and lower surfaces of the dielectric plate are cut to create two additional radiating fins, each shaped by elliptical arcs, as illustrated in Fig. \ref{fig:AntipodalVivaldiAntenna_geometry}.

\begin{figure}[tbp]
    \centering
    \subfloat[]{
        \includegraphics[width=1.0\linewidth]{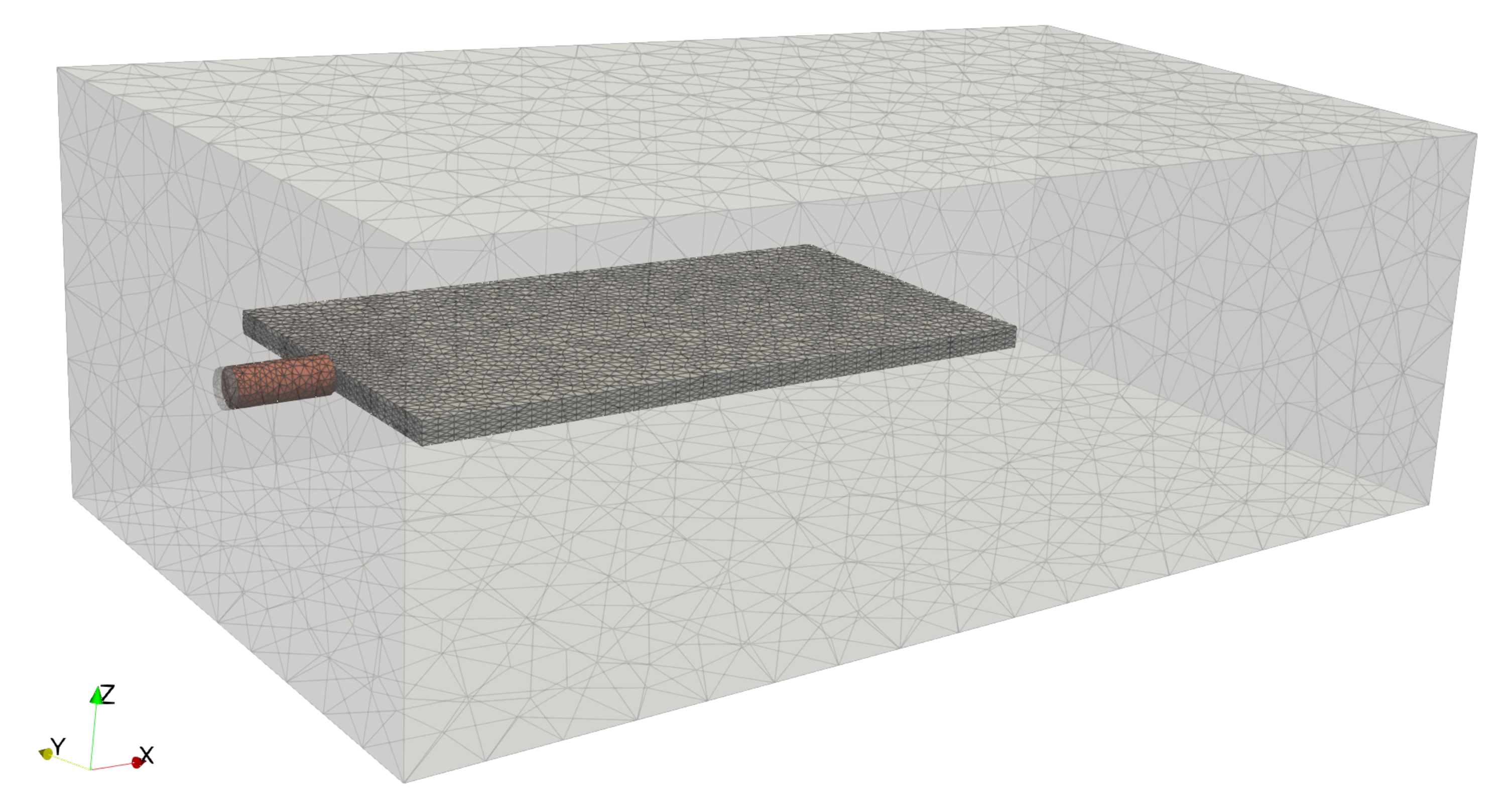}
        \label{fig:AntipodalVivaldiAntenna_mesh_full}
    } \qquad \qquad \qquad
    \subfloat[]{
        \includegraphics[width=1.0\linewidth]{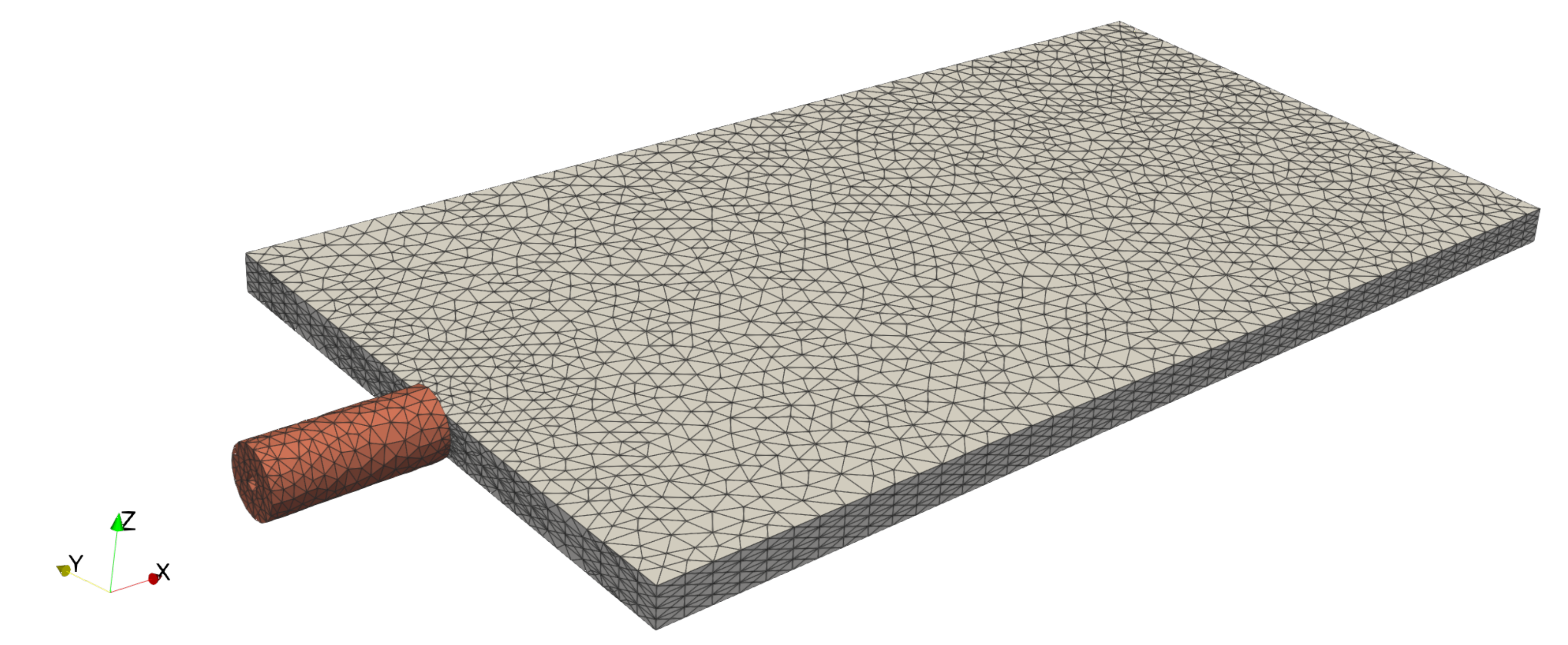}
        \label{fig:AntipodalVivaldiAntenna_mesh_detail}
    }
    \caption{Antipodal Vivaldi antenna mesh. (a) Entire computational domain. (b) Detail of the antenna and the coaxial line.} 
    \label{fig:AntipodalVivaldiAntenna_mesh}
\end{figure}

The numerical simulation considers a stripline-coaxial line transition. The coaxial line has a characteristic impedance of $50 \Omega$, an inner radius of $0.65$ mm, and an outer radius of $2.65$ mm, with the conductor extending $12$ mm outward from the dielectric plate. The surrounding air region, which includes both the antenna and the coaxial line, is modeled as a rectangular box measuring $100$ mm in width, $60$ mm in height, and $172$ mm in length, positioned $24.6$ mm from the dielectric plate. The computational domain considered for both the time and frequency domain simulations is illustrated in Fig.~\ref{fig:AntipodalVivaldiAntenna_mesh}. Specifically, Fig. \ref{fig:AntipodalVivaldiAntenna_mesh_full} depicts the entire domain, including the mesh of the surrounding air region, while Fig. \ref{fig:AntipodalVivaldiAntenna_mesh_detail} provides a detailed view of the antenna and the coaxial line meshes. For wideband analysis, a frequency range from $1$ GHz to $6$ GHz is considered, using as input excitation a Gaussian pulse with a width of $0.2$ ns and a center frequency of $3.5$ GHz. The time domain simulation performed using the time-adaptive ROMTD approach solves the time interval $(0, 120]$ ns with a time step size of $0.005$ ns and initial conditions of zero electromagnetic field. The analysis considers dielectric materials that are lossless and non-magnetic. The relative permittivity is $2.32$ for the dielectric plate and $2.84$ for the coaxial line. To account for radiation losses, absorbing boundary conditions are imposed on the air domain boundaries. Additionally, the radiating fins and the coaxial boundaries are modeled as PEC.

\begin{figure}[tbp]
    \centering
    \includegraphics[width=\linewidth]{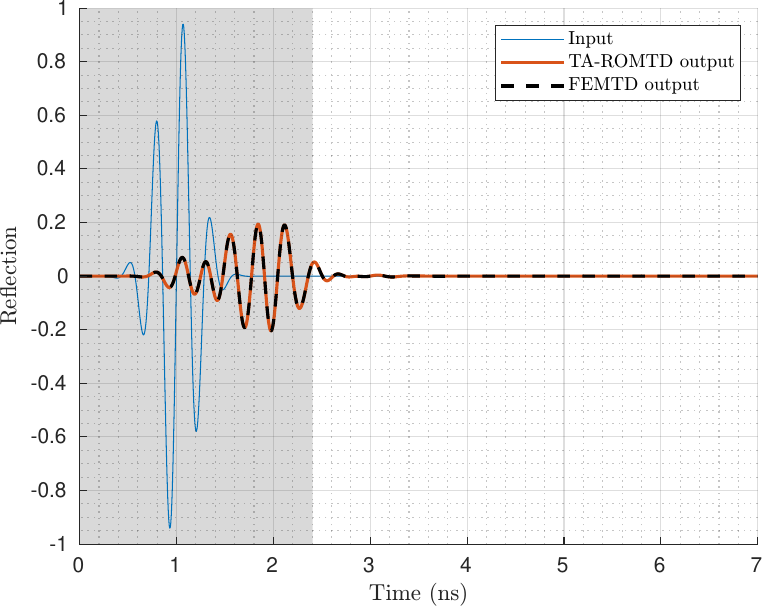}
    \caption{Time domain response of the antipodal Vivaldi antenna from $0$ ns to $7$~ns: the system excitation is represented by the blue curve, the solution obtained using the TA-ROMTD strategy by the orange curve, and the FEMTD reference solution by the black dashed curve. Moreover, the shaded area indicates the time interval solved by the initial FEMTD.}
    \label{fig:AntipodalVivaldiAntenna_TD}
\end{figure}

\begin{figure}[tbp]
    \centering
    \includegraphics[width=\linewidth]{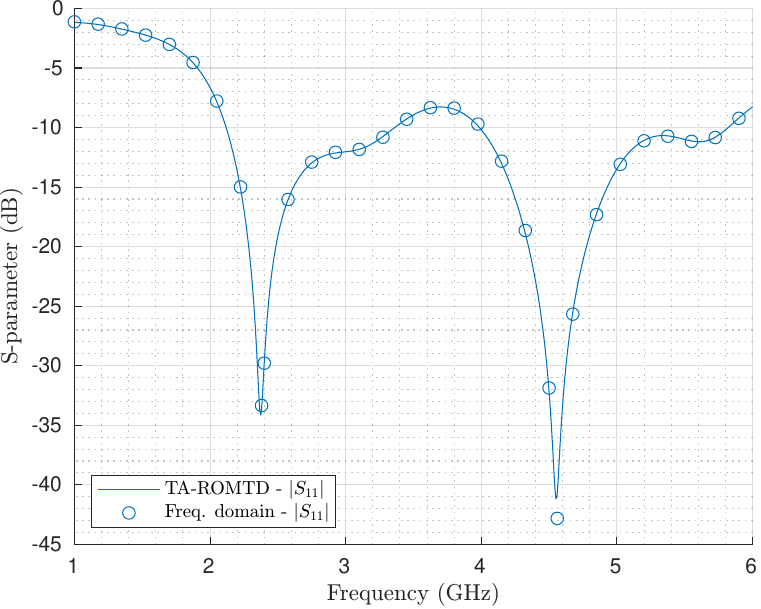}
    \caption{Scattering parameter of the antipodal Vivaldi antenna. The TA-ROMTD results were obtained by applying the FFT to the time domain solution over the interval $(0, 120]$ ns.}
    \label{fig:AntipodalVivaldiAntenna_TDvsFreq}
\end{figure}

\begin{table}[b!]
    \centering
    \caption{Comparison of the time step sizes and DoFs used in TA-ROMTD and FEMTD for time domain analysis of the antipodal Vivaldi antenna.} \label{tab:AntipodalVivaldiAntenna_details}
    \begin{tabular}{lrr}
        \cline{2-3}
        & \textbf{FEMTD}   & \textbf{TA-ROMTD} \\ \hline
        Time step [ns]     & $0.02$          & $0.005$ \\
        No. DoFs           & $511,714$       & $25$ \\ \hline
    \end{tabular}
\end{table}

Fig. \ref{fig:AntipodalVivaldiAntenna_TD} illustrates the scattering parameter response of the antipodal Vivaldi antenna at the coaxial port, 
related to the input return losses over the time interval from $0$ ns to $7$ ns. The figure compares the time domain responses obtained using FEMTD and time-adaptive ROMTD. The system excitation is represented by the blue curve, the time-adaptive ROMTD solution by the orange curve, and the FEMTD reference solution by the black dashed line. Initially, FEMTD solves the first $2.4$ ns using a coarse time step of $0.02$ ns. The generated snapshots are used to construct the ROM basis via POD. Subsequently, ROMTD solves the entire $(0, 120]$ ns interval using a finer $0.005$ ns time step, since the error estimator \eqref{eq:error_estimator} remains below the tolerance of $0.5$ throughout the simulation. As shown, the time-adaptive ROMTD solution closely aligns with the FEMTD results, demonstrating high accuracy and consistency between both approaches.

Table \ref{tab:AntipodalVivaldiAntenna_details} compares the model discretization and time step sizes for both strategies. The FEMTD simulation involves solving a problem with $511,714$ DoFs, whereas the time-adaptive ROMTD reduces this number to just $25$. This reduction in the number of DoFs significantly lowers the computational cost. Specifically, the time-adaptive ROMTD completes the entire $(0, 120]$ ns simulation in approximately $7$ minutes of CPU computations, with $5.5$ minutes dedicated to solving the initial FEMTD interval. In contrast, the FEMTD approach takes over $50$ minutes to solve the first $7$ ns, implying that a full $(0,120]$ ns simulation would take approximately $14$ hours. This demonstrates the efficiency of the time-adaptive ROMTD strategy, which reduces computational time from tens of hours to just a few minutes.

Fig. \ref{fig:AntipodalVivaldiAntenna_TDvsFreq} compares the frequency domain response (circle markers) with the post-processed time domain response (solid line), where the frequency domain response is computed in the $[1, 6]$ GHz frequency band. The results of the two approaches exhibit strong agreement, with only a slight discrepancy at $4.56$ GHz, where the time domain solution slightly overestimates the lower dB values observed in the frequency domain. These results confirm that the time-adaptive ROMTD method accurately captures the response of the electromagnetic system in both the time and frequency domains while significantly reducing computational costs compared to FEMTD analysis.

\subsection{Dielectric Resonator Antenna} \label{sec:results_DRA}

\begin{figure}[tbp]
    \centering
    \includegraphics[width=\linewidth]{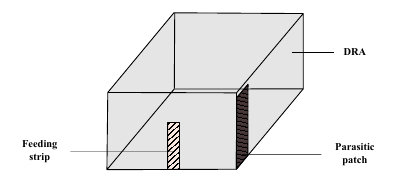}
    \caption{Dielectric resonator antenna geometry representation.}
    \label{fig:DRA_geometry}
\end{figure}

The strip-fed rectangular dielectric resonator antenna (DRA) proposed by \cite{li2005} is analyzed using the described time-adaptive ROMTD. Fig. \ref{fig:DRA_geometry} shows a schematic of the dielectric resonator. It has dimensions of $23.5$ mm in width, $12.34$ mm in height, and $24$ mm in length. A feeding strip, $1$ mm wide and $10$ mm long, is fixed to the center of one side wall of the DRA. Additionally, a parasitic patch, $1$ mm wide and $12$ mm long, is attached to another side wall of the DRA.

\begin{figure}[tbp]
    \centering
    \subfloat[]{
        \includegraphics[width=1.0\linewidth]{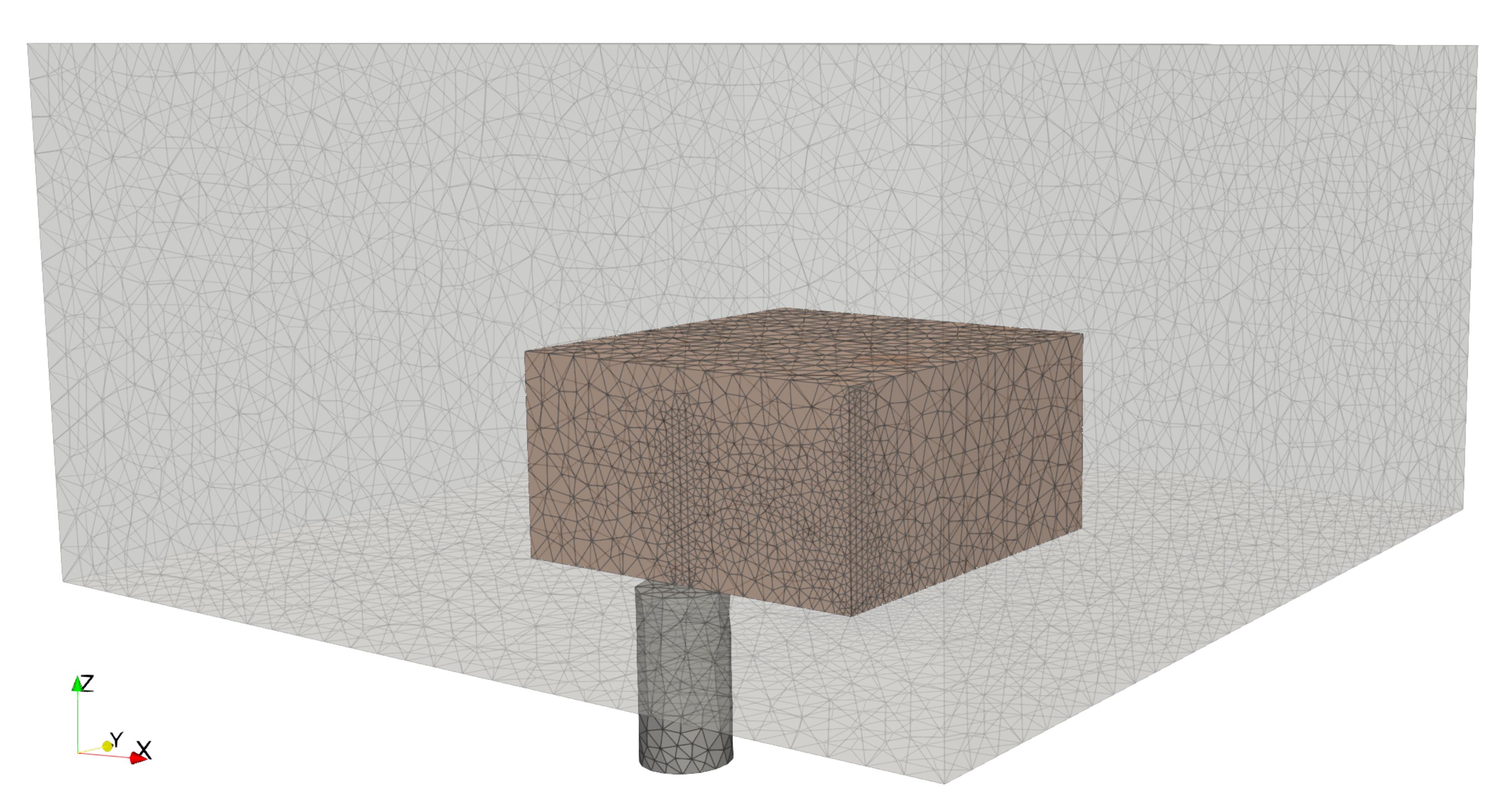}
        \label{fig:DRA_mesh_full}
    } \qquad \qquad \qquad
    \subfloat[]{
        \includegraphics[width=0.9\linewidth]{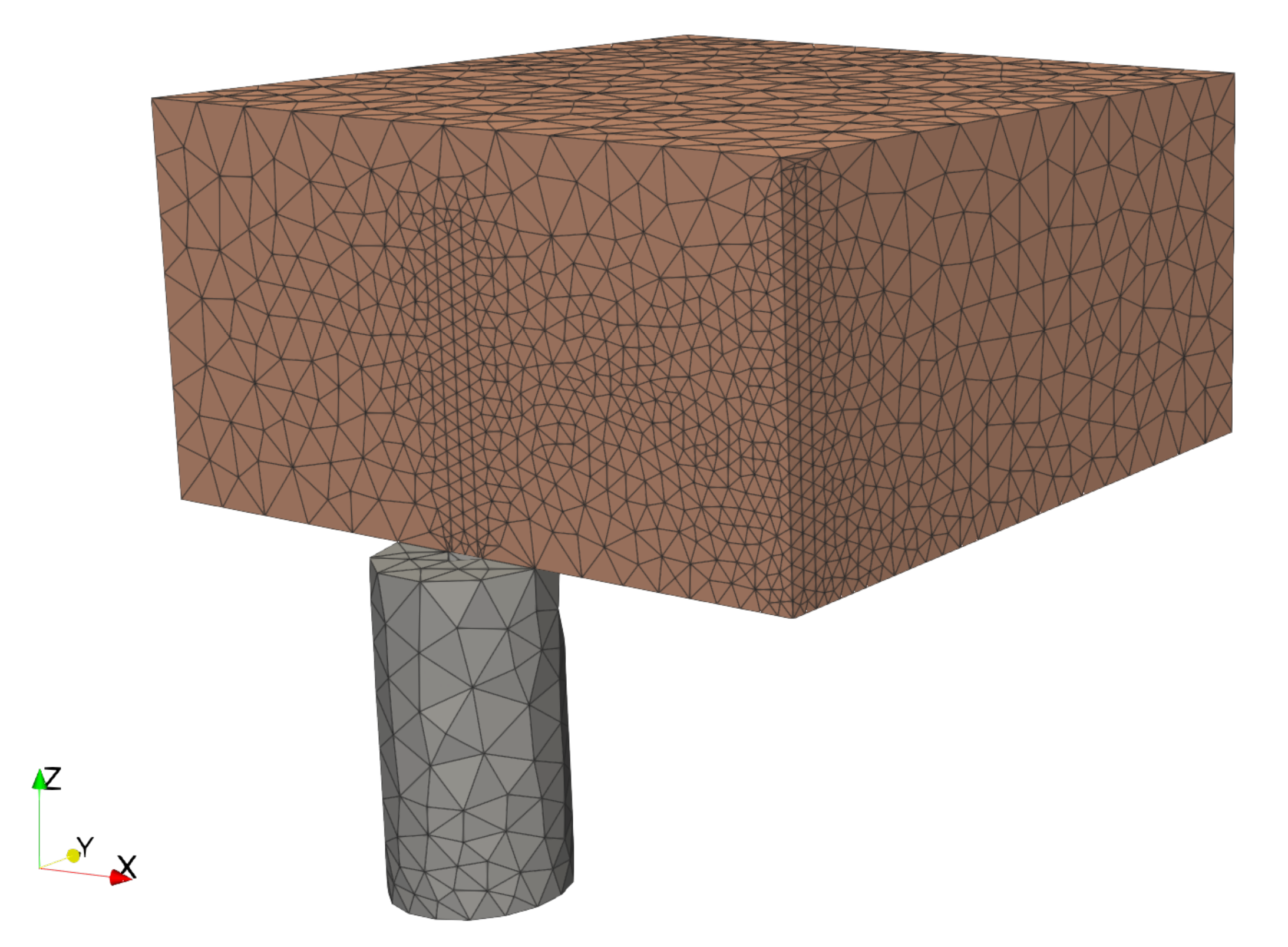}
        \label{fig:DRA_mesh_detail}
    }
    \caption{Dielectric resonator antenna mesh. (a) Entire computational domain. (b) Detail of the DRA.} 
    \label{fig:DRA_mesh}
\end{figure}

In the numerical simulation, the surrounding air region that encompasses the resonator is modeled as a box with dimensions of $60$ mm in width, $30$ mm in height, and $60$ mm in length. The input consists of a $50 \Omega$ coaxial line with a $0.65$ mm inner radius, and a $2.65$ mm outer radius, with the conductor extending $10$ mm in length. Fig. \ref{fig:DRA_mesh} shows the computational domain. In particular, Fig. \ref{fig:DRA_mesh_full} displays the entire computational domain, including the mesh of the surrounding air region, while Fig. \ref{fig:DRA_mesh_detail} focuses on the mesh of the dielectric resonator and the coaxial line. For wideband analysis, the frequency range from $2.5$ GHz to $4.5$ GHz is examined using as input excitation a Gaussian pulse with a width of $0.5$ ns and a center frequency of $3.5$ GHz. The time domain simulation using the time-adaptive ROMTD solves the time interval $(0, 100]$ ns with a time step of $0.001$ ns, starting with zero electromagnetic field as initial conditions. The analysis considers dielectric materials that are lossless and non-magnetic. The relative permittivity is $9.5$ for the dielectric resonator and $2.84$ for the coaxial line. To address radiation losses, absorbing boundary conditions are imposed at the open air domain boundaries. In addition, the feeding strip and the parasitic patch are modeled as PEC, as well as the ground plane supporting the DRA.

\begin{figure}[tbp]
    \centering
    \includegraphics[width=\linewidth]{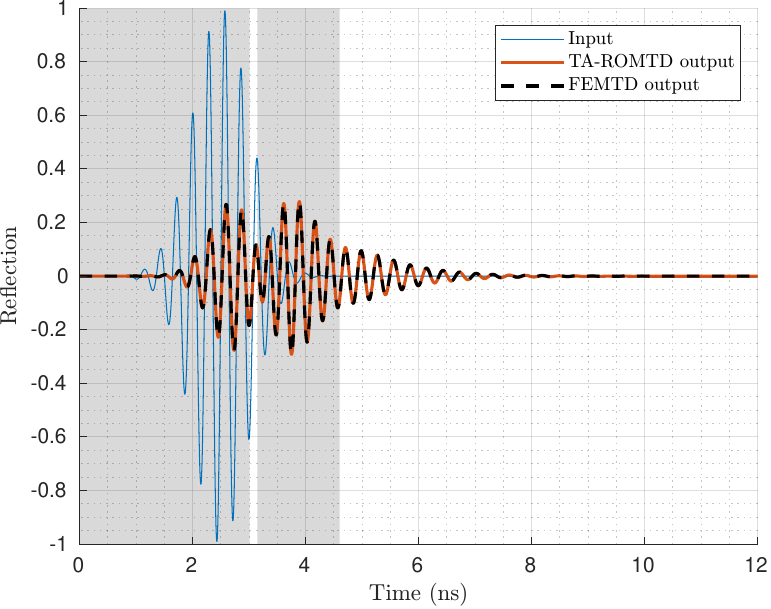}
    \caption{Time domain response of the DRA from $0$ ns to $12$ ns: the system excitation is represented by the blue curve, the solution obtained using the TA-ROMTD strategy by the orange curve, and the FEMTD reference solution by the black dashed curve. Moreover, the shaded areas indicate the time intervals solved using FEMTD.}
    \label{fig:DRA_TD}
\end{figure}

\begin{figure}[tbp]
    \centering
    \includegraphics[width=\linewidth]{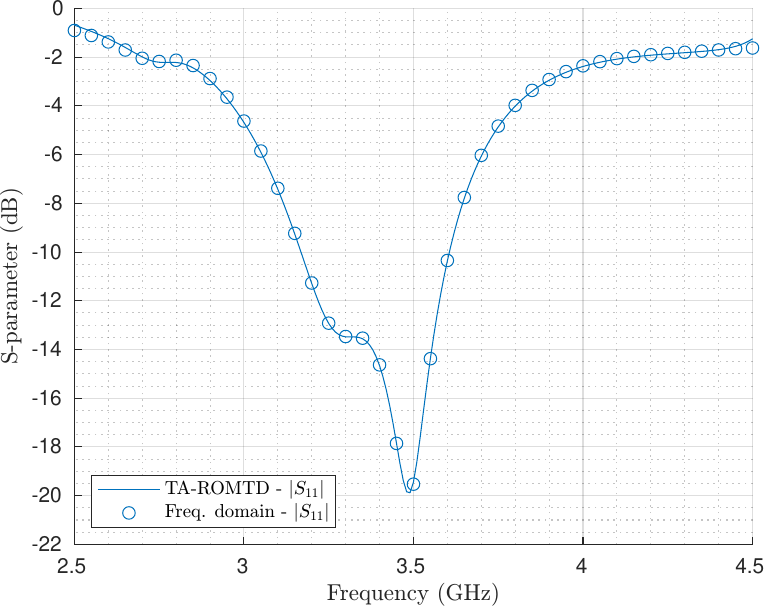}
    \caption{Scattering parameter of the DRA. The TA-ROMTD results were obtained by applying the FFT to the time domain solution over the interval $(0, 100]$ ns.}
    \label{fig:DRA_TDvsFreq}
\end{figure}

\begin{table}[b]
    \centering
    \caption{Comparison of the time step sizes and DoFs used in TA-ROMTD and FEMTD for time domain analysis of the DRA.} \label{tab:DRA_details}
    \begin{tabular}{lrr}
        \cline{2-3}
        & \textbf{FEMTD}   & \textbf{Time-adaptive ROMTD} \\ \hline
        Time step [ns]     & $0.05$      & $0.001$ \\
        No. DoFs           & $484,306$   & $12$ - $42$ \\ \hline
    \end{tabular}
\end{table}

Fig. \ref{fig:DRA_TD} illustrates the scattering parameter response of the DRA, related to input return losses over time from $0$ ns to $12$ ns. The figure compares the time domain responses obtained using both FEMTD and time-adaptive ROMTD. The blue curve represents the system excitation, the orange curve corresponds to the time-adaptive ROMTD solution, and the black dashed line denotes the FEMTD reference solution. In the time-adaptive ROMTD approach, FEMTD is initially used to solve the problem over the first $3$ ns using a large time step of $0.05$ ns. The resulting snapshots are then used to construct the ROM basis via POD. Then, ROMTD solves the simulation from $(0, 3.128]$ ns with a finer time step of $0.001$ ns, at which point the error estimator \eqref{eq:error_estimator} exceeds the set tolerance of $0.05$. The model then switches back to FEMTD to solve the interval $[3.15, 4.6]$ ns, collecting new snapshots to update the ROM basis. Finally, ROMTD solves the remaining time interval from $[3.129, 100]$ ns. The shaded regions in Fig. \ref{fig:DRA_TD} indicate the FEMTD intervals computed with the coarse time step. As shown, the time-adaptive ROMTD solution closely agrees with the FEMTD results, demonstrating high accuracy.

Table \ref{tab:DRA_details} compares the model discretization and time step sizes for both strategies. The FEMTD simulation solves a problem with $484,306$ DoFs, whereas the two ROMs reduce this to $12$ in the first ROMTD and to $42$ in the second. Furthermore, the time-adaptive ROMTD completes the entire simulation of the time interval $(0,100]$ ns in just under $10$ minutes, with $5.5$ minutes spent on the resolution of the FEMTD intervals. In contrast, the FEMTD approach requires over $6$ hours to calculate only the first $12$ ns of the simulation, suggesting that a complete simulation of the time interval $(0,100]$ ns would take approximately $52$ hours.

Furthermore, Fig. \ref{fig:DRA_TDvsFreq} compares the frequency domain response (circle markers) with the post-processed time domain response (solid line). The frequency domain response is obtained by solving the problem in the $[2.5, 4.5]$ GHz frequency band. The results demonstrate a strong agreement between the two approaches, showing that the time domain solution achieves high accuracy at a significantly lower computational cost, reducing the simulation time from tens of hours to just a few minutes.

\subsection{Fully Metallic Dual-Polarization Frequency-Selective Surface} \label{sec:results_Molero_FSS_DualPolarization}
\begin{figure}[b]
    \centering
    \includegraphics[width=0.6\linewidth]{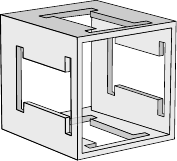}
    \caption{Example of the geometry of a unit cell of the metallic dual-polarization FSS.}
    \label{fig:Molero_geometry}
\end{figure}

The proposed time-adaptive ROMTD strategy is applied to analyze the electromagnetic behavior of the fully metallic dual-polarization frequency selective surface (FSS) designed in \cite{parellada2024}. An example unit cell is depicted in Fig. \ref{fig:Molero_geometry}. The structure consists of seven cells arranged in a repeating pattern of square waveguides, with dog-bone-shaped resonators embedded in the walls, simulating the behavior of a multilayer frequency selective surface. The dimensions of the resonator and the spacing between them vary throughout the structure, with full details provided in \cite{parellada2024data}.

\begin{figure*}[tbp]
    \centering
    \includegraphics[width=\textwidth]{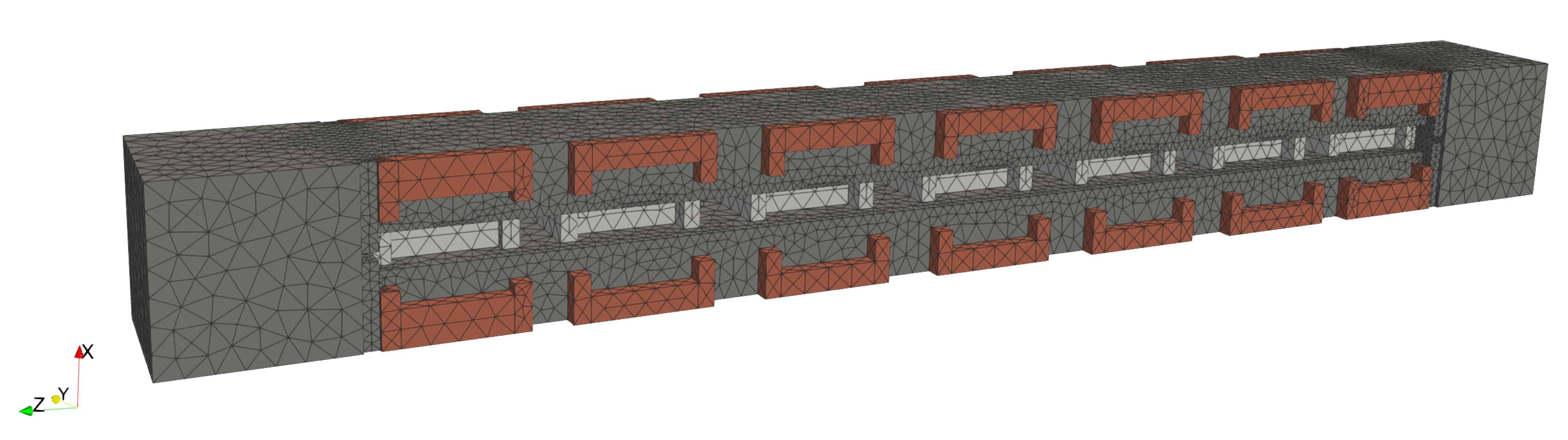}
    \caption{Computational domain of the metallic dual-polarization FSS.}
    \label{fig:Molero_mesh}

    \centering
    \subfloat[]{
        \includegraphics[width=\textwidth]{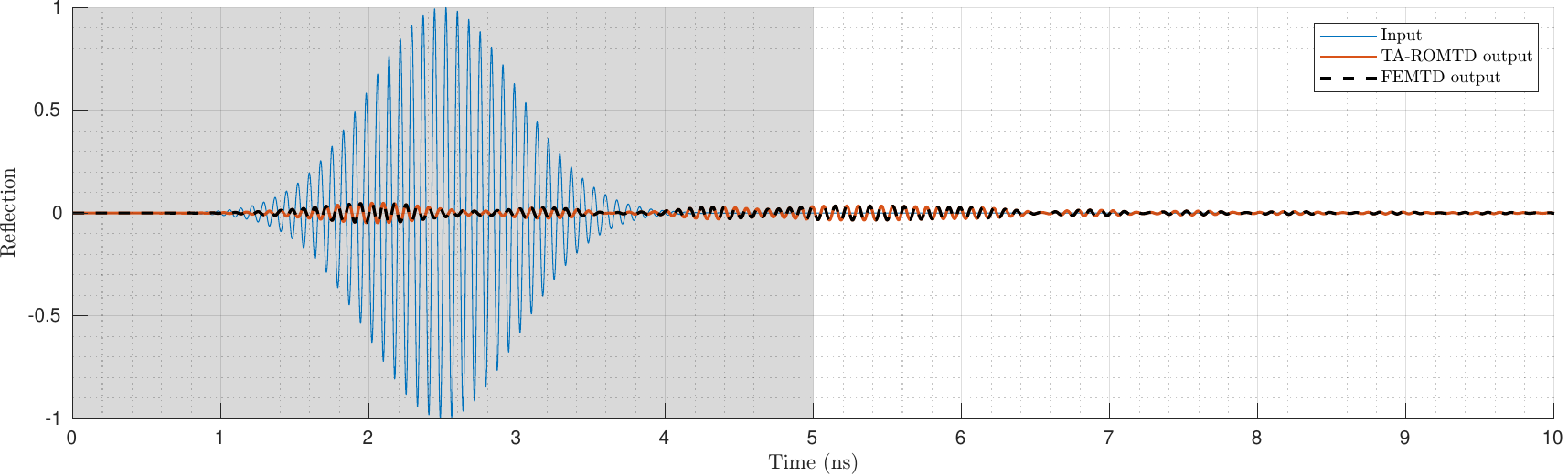}
        \label{fig:Molero_FSS_DualPolarization_TD_reflection}
    } \quad
    \subfloat[]{
        \includegraphics[width=\textwidth]{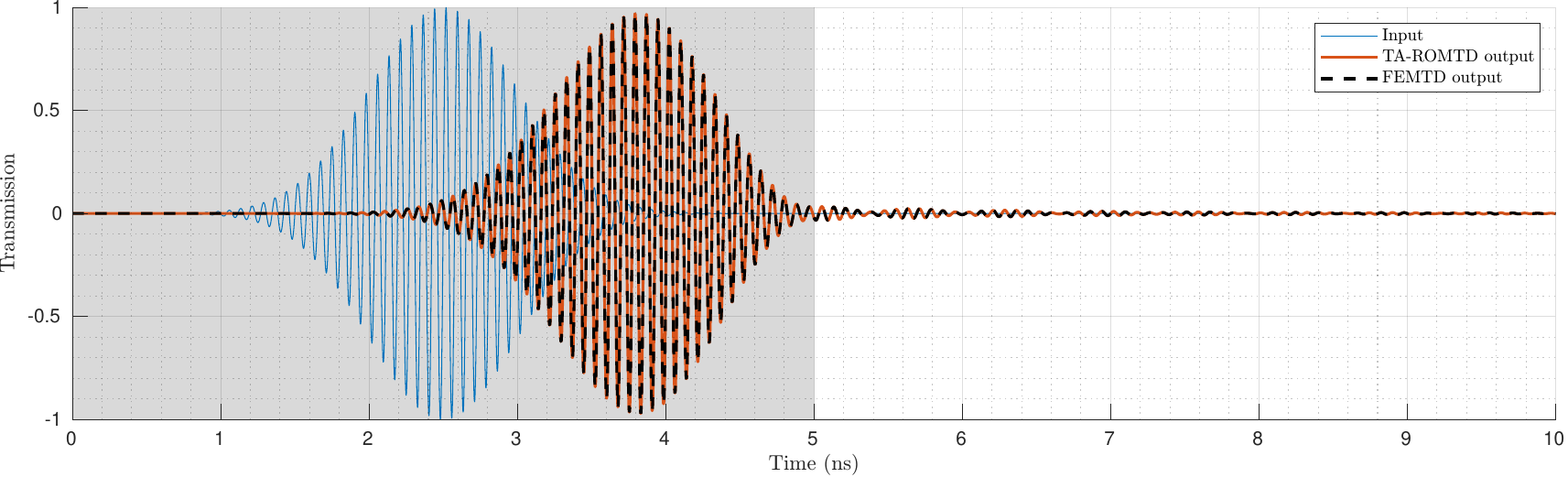}
        \label{fig:Molero_FSS_DualPolarization_TD_transmission}
    }
    \caption{Time domain response of the metallic dual-polarization FSS from $0$ ns to $10$ ns: the system excitation is represented by the blue curve, the solution obtained using the TA-ROMTD strategy by the orange curve, and the FEMTD reference solution by the black dashed curve. Moreover, the shaded area indicates the time interval solved by the initial FEMTD. Subfigures illustrate: (a) Output reflection, and (b) Output transmission.}
    \label{fig:Molero_FSS_DualPolarization_TD}
\end{figure*}

\begin{figure}[tbp]
    \centering
    \includegraphics[width=\linewidth]{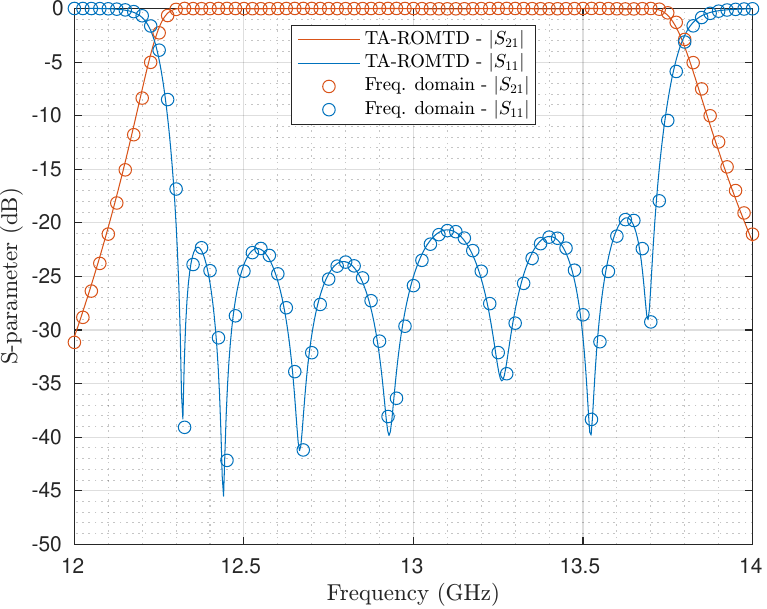}
    \caption{Scattering parameters of the metallic dual-polarization FSS. The TA-ROMTD results were obtained by applying the FFT to the time domain solution over the interval $(0, 250]$ ns.}
    \label{fig:Molero_FSS_DualPolarization_TDvsFreq}
\end{figure}

\begin{table}[b]
    \centering
    \caption{Comparison of the time step sizes and DoFs used in TA-ROMTD and FEMTD for time domain analysis of the metallic dual-polarization FSS.} \label{tab:Molero_FSS_DualPolarization_details}
    \begin{tabular}{lrr}
        \cline{2-3}
        & \textbf{FEMTD}   & \textbf{Time-adaptive ROMTD} \\ \hline
        Time step [ns]     & $0.005$        & $0.001$ \\
        No. DoFs           & $285,940$      & $16$    \\ \hline
    \end{tabular}
\end{table}

Fig. \ref{fig:Molero_mesh} shows the computational domain used for the simulations in the time and frequency domains. The frequency range from $12$ GHz to $14$ GHz is analyzed using a Gaussian pulse of $0.5$ ns width and centered at $13$ GHz as input excitation. The time domain simulation using the time-adaptive ROMTD solves the time interval $(0, 250]$ ns with a time step of $0.001$ ns, starting with zero electromagnetic field as initial conditions. All subdomains within the computational domain are treated as vacuum, with PEC and PMC boundary conditions used to simulate a periodic environment.

The time domain response of the metallic dual-polarization FSS, from $0$ ns to $10$ ns, is shown in Fig. \ref{fig:Molero_FSS_DualPolarization_TD}. The figure compares FEMTD (black dashed line) and time-adaptive ROMTD (orange curve) with the system excitation (blue curve). In the time-adaptive ROMTD, FEMTD solves the first $5$ ns using a $0.005$ ns step size. These results are used to construct a ROM basis via POD. ROMTD then solves the entire $(0, 250]$ ns interval with a $0.001$ ns time step, since the error estimator \eqref{eq:error_estimator} remains below the tolerance of 0.05 throughout. Figs. \ref{fig:Molero_FSS_DualPolarization_TD_reflection} and \ref{fig:Molero_FSS_DualPolarization_TD_transmission} illustrate the output reflection and transmission responses, respectively, over the time period $(0,10]$ ns, with the shaded region indicating the initial FEMTD interval. As shown, the time-adaptive ROMTD solution closely matches the FEMTD results, demonstrating high accuracy and consistency between both approaches.

Table \ref{tab:Molero_FSS_DualPolarization_details} compares the model discretization and time step sizes for both strategies. The FEMTD simulation considers $285,940$ DoFs, whereas the time-adaptive ROMTD reduces this to just $16$. The full simulation of $(0,250]$ ns is completed in $45$ minutes using the time-adaptive ROMTD, with $31$ minutes required to solve the FEMTD interval. In contrast, FEMTD alone requires $4.5$ hours to solve the first $10$ ns, implying that a simulation of $250$ ns would take approximately $113$ hours. Thus, the time-adaptive ROMTD approach reduces the computational cost by three orders of magnitude.

Finally, Fig. \ref{fig:Molero_FSS_DualPolarization_TDvsFreq} compares the frequency domain response (circle markers) with the post-processed time domain response (solid line). The frequency domain response is obtained by solving the problem in the range from $12$ GHz to $14$ GHz. The strong agreement between both approaches confirms that the time-adaptive ROMTD maintains high accuracy while drastically reducing the computational cost.

\subsection{Mushroom-type Electromagnetic Bandgap Structure} \label{sec:results_quevedo3}
This section uses the time-adaptive ROMTD methodology to analyze a mushroom-type electromagnetic bandgap (EBG) surface embedded in a dielectric parallel plate, proposed in \cite{mouris2020}. The structure consists of six metallic patches, each $7$ mm wide and $3.5$ mm long, separated by $1$ mm gaps. Additionally, each patch includes an edge via measuring $1.524$ mm in width and $0.3$ mm in length. The parallel plate has dimensions of $8$ mm in width, $3.048$ mm in height, and $33$ mm in length, with the patches centered within the structure.

\begin{figure}[tbp]
    \centering
    \subfloat[]{
        \includegraphics[width=1.0\linewidth]{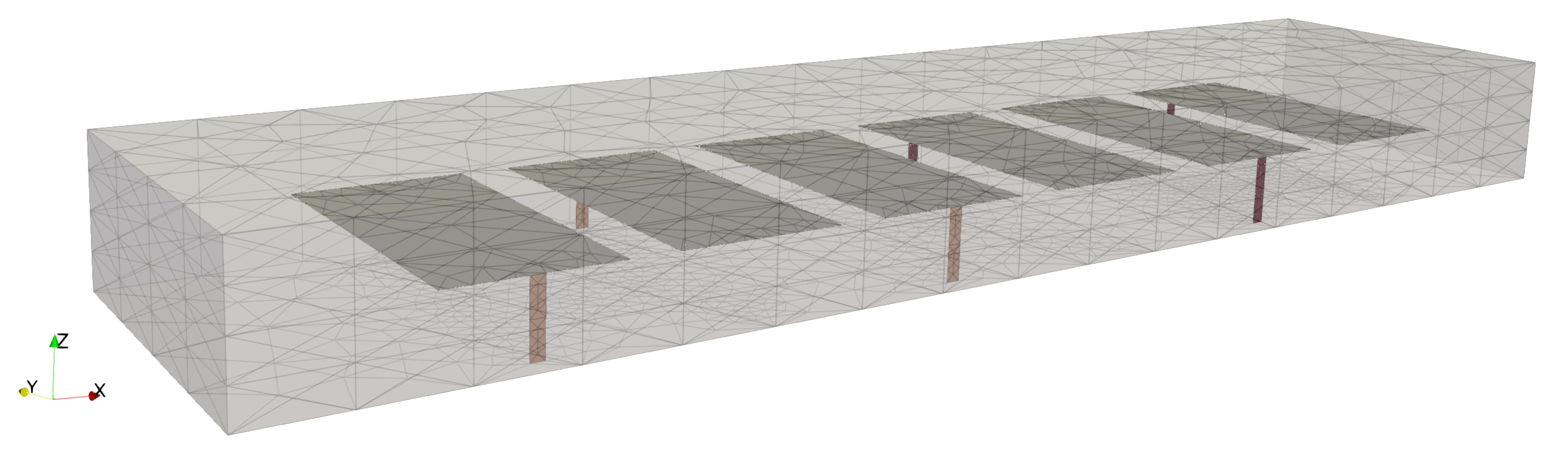}
        \label{fig:Quevedo3_mesh_full}
    } \qquad \qquad \qquad
    \subfloat[]{
        \includegraphics[width=1.0\linewidth]{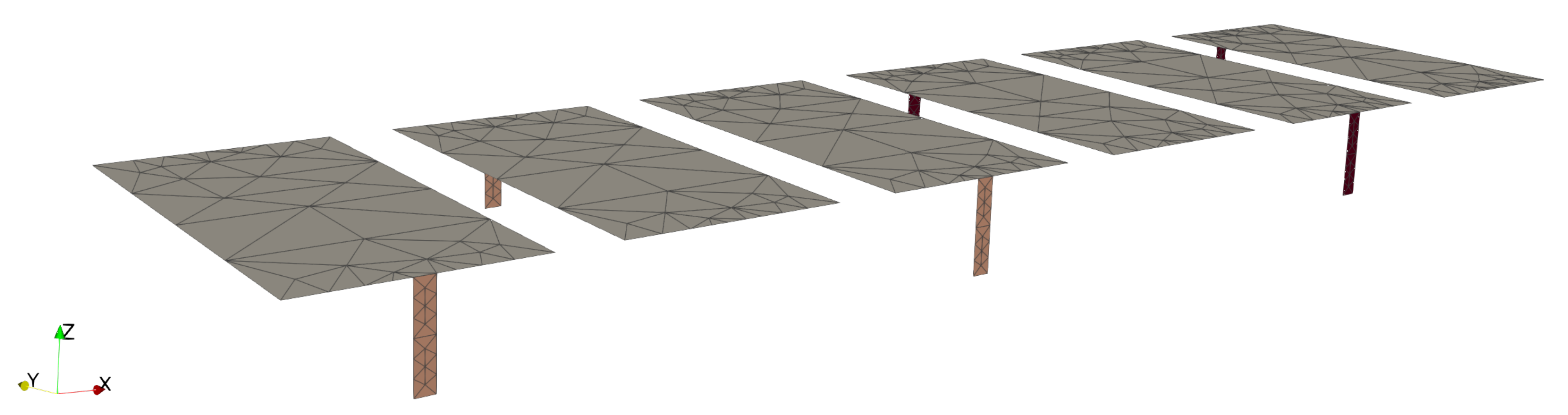}
        \label{fig:Quevedo3_mesh_PEC}
    }
    \caption{Parallel-plate with a mushroom-type EBG surface mesh. (a) Entire computational domain. (b) Detail of the metallic patches.}
    \label{fig:Quevedo3_mesh}
\end{figure}

Fig. \ref{fig:Quevedo3_mesh} illustrates the computational domain considered for simulations in both time and frequency domains. The $[1,5]$ GHz frequency band is analyzed using a Gaussian pulse centered at $3$ GHz with a width of $0.25$ ns. The time domain simulation runs from $(0, 500]$ ns with a time step of $0.001$ ns. The analysis assumes lossless and non-magnetic material with a relative permittivity of $3.48$. Metallic patches and edge vias are modeled as PEC, while the air box boundaries at the maximum and minimum values of the $y$-coordinate are treated as PMC.

\begin{figure}[tbp]
    \centering
    \includegraphics[width=\linewidth]{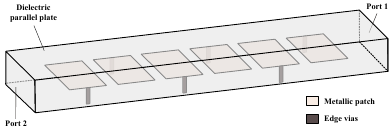}
    \caption{Geometry of the mushroom-type EBG.}
    \label{fig:Quevedo3_geometry}
\end{figure}

\begin{figure*}[tbp]
    \centering
    \subfloat[]{
        \includegraphics[width=\textwidth]{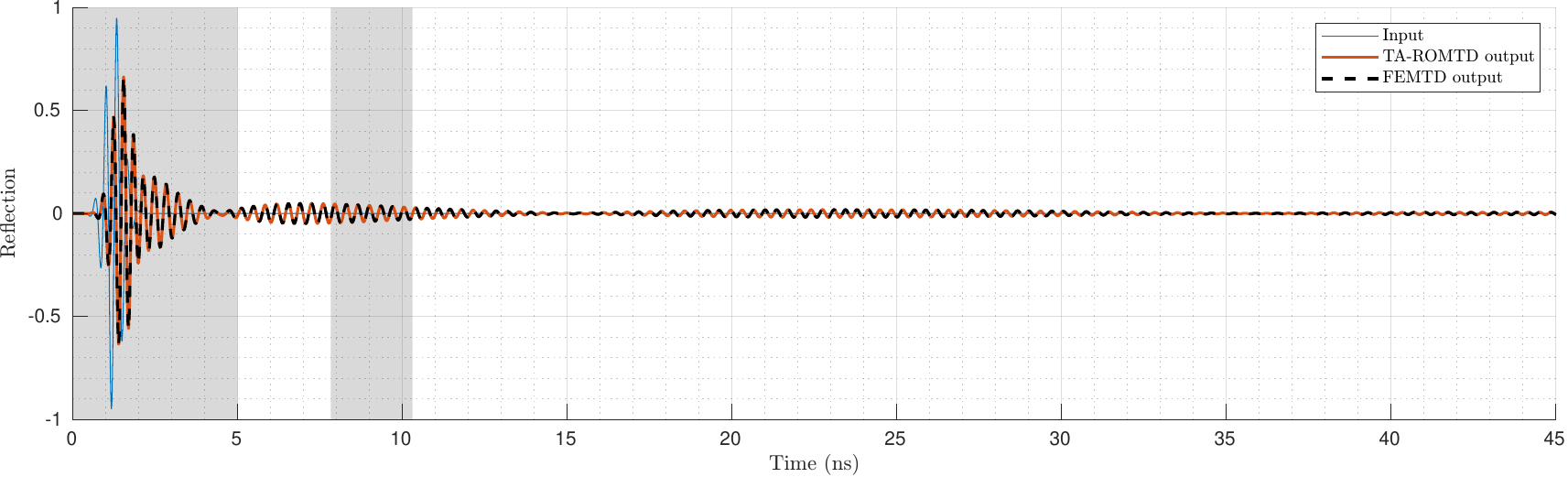}
        \label{fig:Quevedo_ParallelPlate_TD_reflection}
    } \quad
    \subfloat[]{
        \includegraphics[width=\textwidth]{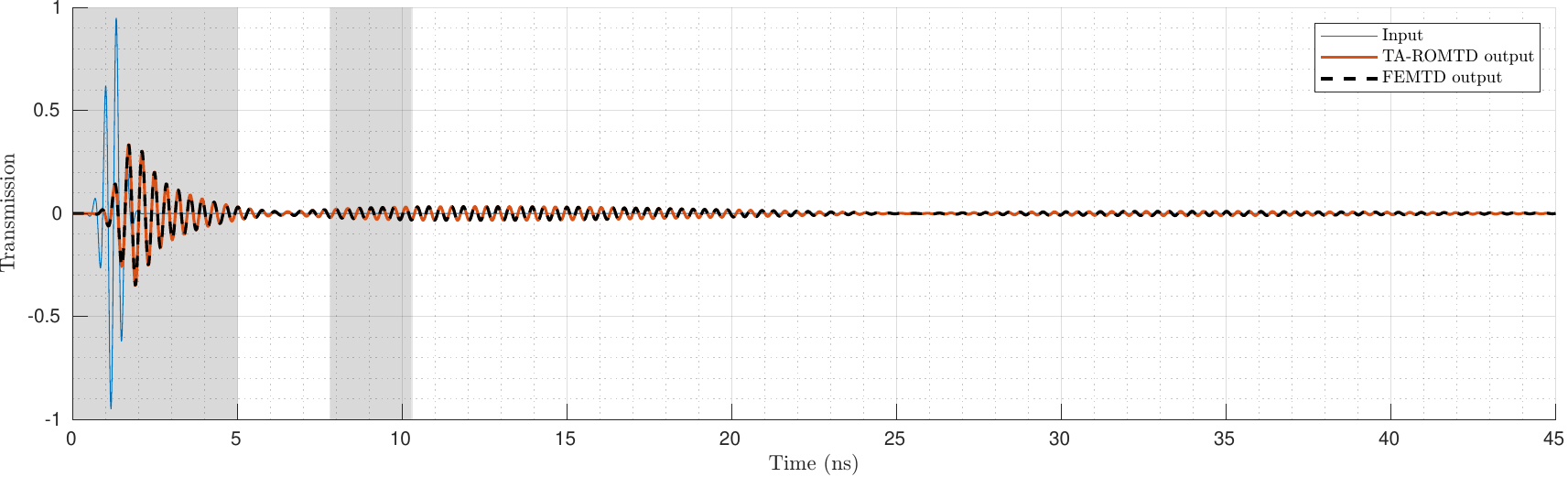}
        \label{fig:Quevedo_ParallelPlate_TD_transmission}
    }
    \caption{Time domain response of the parallel-plate waveguide with a mushroom-type EBG surface from $0$ ns to $45$ ns: the system excitation is represented by the blue curve, the solution obtained using the TA-ROMTD strategy by the orange curve, and the FEMTD reference solution by the black dashed curve. Moreover, the shaded areas indicate the time intervals solved using FEMTD. Subfigures illustrate: (a) Output reflection, and (b) Output transmission.}
    \label{fig:Quevedo_ParallelPlate_TD}

    \centering
    \subfloat[]{
    \includegraphics[width=0.47\linewidth]{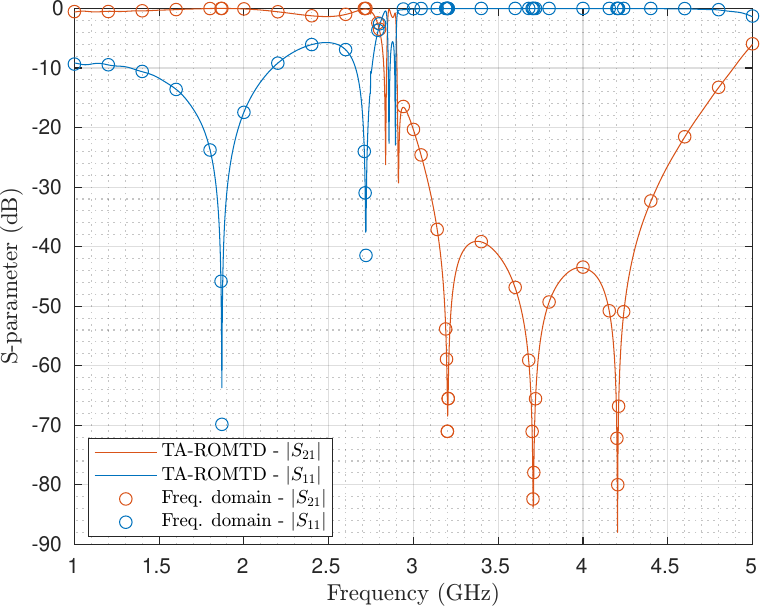}
        \label{fig:Quevedo_ParallelPlate_TDvsFreq_full}
    } \quad
    \subfloat[]{
    \includegraphics[width=0.47\linewidth]{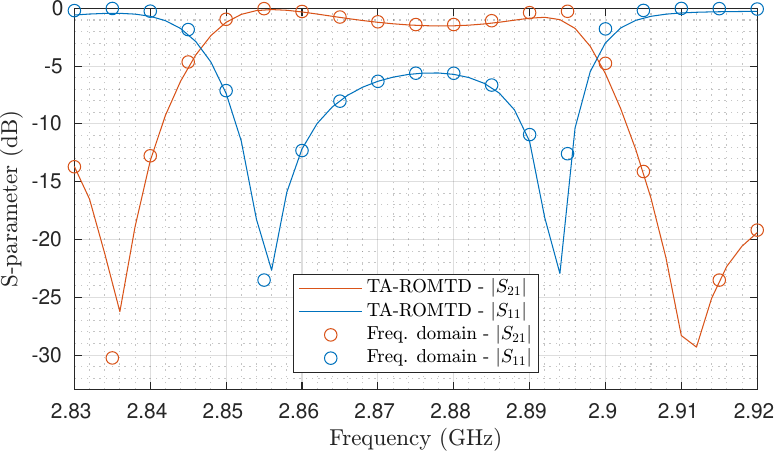}
        \label{fig:Quevedo_ParallelPlate_TDvsFreq_zoom}
    }
    \caption{Scattering parameters of the parallel-plate waveguide with a mushroom-type EBG surface. The ROMTD results were obtained by applying the FFT to the time domain solution over the interval $(0, 500]$ ns. Subfigures illustrate: (a) Entire frequency range, and (b) Frequencies between $2.83$ GHz and $2.92$ GHz.}
    \label{fig:Quevedo_ParallelPlate_TDvsFreq}
\end{figure*}

\begin{table}[tbp]
    \centering
    \caption{Comparison of the time step sizes and DoFs used in TA-ROMTD and FEMTD for time domain analysis of the parallel-plate waveguide with a mushroom-type EBG surface.} \label{tab:Quevedo3_details}
    \begin{tabular}{lrr}
        \cline{2-3}
        & \textbf{FEMTD}   & \textbf{Time-adaptive ROMTD} \\ \hline
        Time step [ns]     & $0.025$        & $0.001$ \\
        No. DoFs           & $84,320$       & $19$ - $93$ \\ \hline
    \end{tabular}
\end{table}

The time domain response of the parallel-plate waveguide with the mushroom-type EBG surface, from $0$ ns to $45$ ns, is depicted in Fig. \ref{fig:Quevedo_ParallelPlate_TD}. This figure compares FEMTD (black dashed line) and time-adaptive ROMTD (orange curve) solutions against system excitation (blue curve). In the time-adaptive ROMTD approach, FEMTD is initially used to solve the problem over the first $5$ ns using a large time step of $0.025$ ns and assuming a zero electromagnetic field as initial conditions. These results generate the ROM basis via POD. ROMTD then solves the simulation from $0$ ns to $7.806$ ns with time step $0.001$ ns. At which point the error estimator \eqref{eq:error_estimator} surpasses the given tolerance of $0.05$. Then, the model switches back to FEMTD for the $[7.825, 10.3]$ ns time interval, collecting new snapshots to update the ROM basis. Finally, ROMTD completes the remaining $[7.807, 500]$ ns time interval using the last two time steps of the previous ROMTD interval as initial conditions. The shaded regions in Figs.~ \ref{fig:Quevedo_ParallelPlate_TD_reflection} and \ref{fig:Quevedo_ParallelPlate_TD_transmission} highlight the FEMTD intervals. As shown, the time-adaptive ROMTD solution closely follows the FEMTD results, confirming the high accuracy of this strategy.

Table \ref{tab:Quevedo3_details} compares the model discretization and time step sizes for both strategies. The FEMTD considers $84,320$ DoFs, whereas ROMTD reduces this to $19$ in the first ROMTD and $93$ in the second. The full $(0,500]$ ns simulation using the time-adaptive ROMTD is completed in $22$ minutes, with $3$ minutes spent on the resolution of the FEMTD intervals. In this example, the cost of resolution with the reduced model is slightly higher than in the previous examples because the second ROMTD has a dimension of $93$. On the other hand, FEMTD alone requires $6.3$ hours to solve just $45$ ns, which implies that a simulation of $500$ ns would take approximately $70$ hours.

Finally, Fig. \ref{fig:Quevedo_ParallelPlate_TDvsFreq} compares the frequency domain response (circle markers) with the post-processed time domain response (solid line) from $1$ GHz to $5$ GHz. A detailed view of the results between $2.83$ GHz and $2.92$ GHz is provided in Fig.~\ref{fig:Quevedo_ParallelPlate_TDvsFreq_zoom}. The comparison shows a strong agreement between both approaches, with only minor discrepancies, while significantly reducing the simulation time from several hours to just minutes.

\section{Conclusions} \label{sec:conclusions}
A novel time-adaptive reduced order model (TA-ROMTD) has been proposed for electromagnetic simulations in the time domain using FEMTD. This approach significantly improves computational efficiency while maintaining the accuracy of the solution. It automatically switches between FEMTD and ROMTD based on the evaluation of an error estimator, ensuring an accurate ROM solution without requiring prior knowledge of the problem. Using a coarse time step during FEMTD time intervals and reducing the number of DoFs for most of the simulation, TA-ROMTD drastically reduces computational costs.

The efficiency of the TA-ROMTD strategy has been demonstrated in various electromagnetic applications, including an antipodal Vivaldi antenna, a dielectric resonator antenna, a fully metallic dual-polarization frequency-selective surface, and a mushroom-type electromagnetic bandgap structure. This method achieves substantial computational savings, reducing simulation time by orders of magnitude compared to FEMTD. Specifically, simulations that took tens of hours with FEMTD were completed in minutes using TA-ROMTD. Additionally, the proposed approach maintains high accuracy in the solution, with time domain results closely matching FEMTD solutions.

TA-ROMTD provides a powerful tool for accelerating time domain electromagnetic simulations, enabling large-scale and broadband analyses for antenna and microwave circuit design. Its successful application to a variety of electromagnetic problems highlights its potential for addressing more complex challenges, such as time-varying media properties \cite{koivurova2023,pacheco-pena2024} or nonlinear problems.

\appendices
\section{Proper orthogonal decomposition} \label{sec:POD}

The \emph{proper orthogonal decomposition} (POD) \cite{benner2015,brunton2019} is a widely-used technique for generating an orthonormal basis that reduces the dimensionality of complex, high-dimensional problems, such as the one described in Section \ref{sec:problem_statement}. By applying POD, an orthogonal basis can be derived that captures the essential dynamics reflected in a given set of solutions, providing a more computationally manageable representation of the system \cite{lucia2004, chaturantabut2012}.

Consider electric field solutions or snapshots at $m$ time steps obtained by solving the discrete system \eqref{eq:FEM_time_discretization}
\begin{equation}
    \mathbf{E}^{1}, \mathbf{E}^{2}, \ldots, \mathbf{E}^{m}.
\end{equation}
By using POD, the electric field snapshots can be expressed as:
\begin{equation} \label{eq:POD_E}
    \mathbf{E} (\mathbf{r}, t) = \sum_{j=1}^{m} a_{j}(t) \ \boldsymbol{\psi}_{j} (\mathbf{r}).
\end{equation}
In the previous representation, $\{\boldsymbol{\psi}_{j} (\mathbf{r})\}_{j=1}^{m}$ denotes the set of \emph{POD-modes}, while the coefficients $a_{j} (t)$ indicate the amplitudes of these modes. The POD-modes form an orthonormal basis that captures the most significant features of the selected snapshots, focusing on the dominant dynamics of the system.

For all $i, j = 1, \ldots, m$, the correlation matrix $\mathbf{C}$ is defined as:
\begin{equation}
    C_{ij} = \langle\mathbf{E}^{i}, \mathbf{E}^{j}\rangle,
\end{equation}
where $\langle\cdot,\cdot\rangle$ denotes the inner product between two snapshots. To compute the POD-modes, the following eigenvalue problem must be solved for $l = 1, \ldots, m$:
\begin{equation}
    \mathbf{C} \mathbf{v}_{l} = \lambda_{l} \mathbf{v}_{l},
\end{equation}
where $\mathbf{v}_{l} \in \mathbb{R}^{m}$ is the $l$-th eigenvector of the correlation matrix $\mathbf{C}$, and $\lambda_{l} \in \mathbb{R}$ is the corresponding eigenvalue.

Once the eigenvectors of the correlation matrix are computed, the $l$-th POD-mode is calculated as a linear combination of the collected snapshots using the following expression:
\begin{equation}
    \boldsymbol{\psi}_{l} = \frac{1}{\sqrt{\lambda_{l}}} \mathbf{X} \mathbf{v}_{l},
\end{equation}
where $\mathbf{X}$ is the matrix containing the snapshots as columns:
\begin{equation}
    \begin{bmatrix}
        \vert & \vert & & \vert \\
        \mathbf{E}^{1} & \mathbf{E}^{2} & \ldots & \mathbf{E}^{m}\\
        \vert & \vert & & \vert 
    \end{bmatrix}.
\end{equation}
For a detailed explanation of this computation, the reader is referred to \cite{hesthaven2016}.

The POD decomposition \eqref{eq:POD_E} can be truncated by including only the first $M < m$ modes from the set of snapshots, thus obtaining a reduced basis for the space defined by these snapshots. This approach yields an optimal set of basis vectors that minimizes the mean squared error when approximating snapshots \cite{chaturantabut2012,feng2017}.

\section*{Acknowledgement}
This work has been developed within the framework of the activities of the project \emph{Plasma reconfigUrabLe metaSurface tEchnologies} (PULSE), funded by the European Innovation Council under the EIC Pathfinder Open 2022 program (protocol number 101099313). The project website is: \url{https://www.pulse-pathfinder.eu/}.

\bibliographystyle{sty/IEEEtran}
\bibliography{bibliography/references}

\begin{thebibliography}{10}
\providecommand{\url}[1]{#1}
\csname url@samestyle\endcsname
\providecommand{\newblock}{\relax}
\providecommand{\bibinfo}[2]{#2}
\providecommand{\BIBentrySTDinterwordspacing}{\spaceskip=0pt\relax}
\providecommand{\BIBentryALTinterwordstretchfactor}{4}
\providecommand{\BIBentryALTinterwordspacing}{\spaceskip=\fontdimen2\font plus
\BIBentryALTinterwordstretchfactor\fontdimen3\font minus
  \fontdimen4\font\relax}
\providecommand{\BIBforeignlanguage}[2]{{%
\expandafter\ifx\csname l@#1\endcsname\relax
\typeout{** WARNING: IEEEtran.bst: No hyphenation pattern has been}%
\typeout{** loaded for the language `#1'. Using the pattern for}%
\typeout{** the default language instead.}%
\else
\language=\csname l@#1\endcsname
\fi
#2}}
\providecommand{\BIBdecl}{\relax}
\BIBdecl

\bibitem{tan2007}
E.~L. Tan, ``{Unconditionally Stable LOD-FDTD Method for 3-D Maxwell's
  Equations},'' \emph{{IEEE} Microw. Wirel. Compon. Lett.}, vol.~17, no.~2, pp.
  85--87, 2007.

\bibitem{liu2024}
S.~Liu and E.~L. Tan, ``{Unified Leapfrog CDI-FDTD Methods With Weakly
  Conditional Stability and Hybrid Implicit-Explicit Schemes},'' \emph{{IEEE}
  Trans. Antennas Propag.}, vol.~72, no.~8, pp. 6652--6662, 2024.

\bibitem{taflove2005}
A.~Taflove and S.~C. Hagness, \emph{{Computational Electrodynamics: The
  Finite-Difference Time-Domain Method}}, 3rd~ed.\hskip 1em plus 0.5em minus
  0.4em\relax Artech House, 2015.

\bibitem{kenan2024}
K.~Tekbaş, J.-P. Berenger, L.~M.~D. Angulo, M.~Ruiz-Cabello, and S.~G. Garcia,
  ``{FDTD Voxels-in-Cell Method With Debye Media},'' \emph{{IEEE} Trans.
  Antennas Propag.}, vol.~72, no.~5, pp. 4431--4439, 2024.

\bibitem{valverde2024}
A.~J.~M. Valverde, M.~Ruiz-Cabello, A.~R. Bretones, A.~G. Bravo, and S.~G.
  García, ``{Analysis and Improvement of the Stability of a 3-D FDTD
  Subgridding Method by Applying an LECT-Based Technique},'' \emph{{IEEE}
  Trans. Antennas Propag.}, vol.~72, no.~1, pp. 791--799, 2024.

\bibitem{angulo2015}
L.~D. Angulo, J.~Alvarez, M.~F. Pantoja, S.~G. Garcia, and A.~Bretones,
  ``{Discontinuous Galerkin Time Domain Methods in Computational
  Electrodynamics: State of the Art},'' in \emph{Forum Electromagn. Res.
  Methods Appl. Technol.}, vol.~10, no.~4, 2015.

\bibitem{yan2017}
S.~Yan and J.-M. Jin, ``{A Dynamic $p$-Adaptive DGTD Algorithm for
  Electromagnetic and Multiphysics Simulations},'' \emph{{IEEE} Trans. Antennas
  Propag.}, vol.~65, no.~5, pp. 2446--2459, 2017.

\bibitem{chen2018}
G.~Chen, L.~Zhao, W.~Yu, S.~Yan, K.~Zhang, and J.-M. Jin, ``{A General Scheme
  for the Discontinuous Galerkin Time-Domain Modeling and S-Parameter
  Extraction of Inhomogeneous Waveports},'' \emph{{IEEE} Trans. Microwave
  Theory Tech.}, vol.~66, no.~4, pp. 1701--1712, 2018.

\bibitem{chen2013}
J.~Chen and Q.~H. Liu, ``{Discontinuous Galerkin Time-Domain Methods for
  Multiscale Electromagnetic Simulations: A Review},'' \emph{Proceedings of the
  {IEEE}}, vol. 101, no.~2, pp. 242--254, 2013.

\bibitem{ren2024}
Q.~Ren, H.~Cao, and Q.~H. Liu, ``{Novel Efficient Discontinuous Galerkin Time-
  Domain Modeling of Dispersive Chiral Metamaterials via Media
  Homogenization},'' \emph{{IEEE} Trans. Microwave Theory Tech.}, vol.~72,
  no.~9, pp. 5218--5227, 2024.

\bibitem{dosopoulos2010MAG}
S.~Dosopoulos and J.-F. Lee, ``{Interior Penalty Discontinuous Galerkin Method
  for the Time-Domain Maxwell's Equations},'' \emph{IEEE Trans. Magn.},
  vol.~46, no.~8, pp. 3512--3515, 2010.

\bibitem{dosopoulos2010TAP}
------, ``{Interior Penalty Discontinuous Galerkin Finite Element Method for
  the Time-Dependent First Order Maxwell's Equations},'' \emph{{IEEE} Trans.
  Antennas Propag.}, vol.~58, no.~12, pp. 4085--4090, 2010.

\bibitem{dosopoulos2013}
S.~Dosopoulos, B.~Zhao, and J.-F. Lee, ``{Non-conformal and parallel
  discontinuous Galerkin time domain method for Maxwell's equations: EM
  analysis of IC packages},'' \emph{J. Comput. Phys.}, vol. 238, pp. 48--70,
  2013.

\bibitem{angulo2010}
J.~Alvarez, L.~D. Angulo, M.~F. Pantoja, A.~R. Bretones, and S.~G. Garcia,
  ``{Source and Boundary Implementation in Vector and Scalar DGTD},''
  \emph{{IEEE} Trans. Antennas Propag.}, vol.~58, no.~6, pp. 1997--2003, 2010.

\bibitem{angulo2012}
J.~Alvarez, L.~Angulo, A.~Rubio~Bretones, and S.~G. Garcia, ``{A Spurious-Free
  Discontinuous Galerkin Time-Domain Method for the Accurate Modeling of
  Microwave Filters},'' \emph{{IEEE} Trans. Microw. Theory Tech.}, vol.~60,
  no.~8, pp. 2359--2369, 2012.

\bibitem{lee1997}
J.-F. Lee, R.~Lee, and A.~Cangellaris, ``Time-domain finite-element methods,''
  \emph{{IEEE} Trans. Antennas Propag.}, vol.~45, no.~3, pp. 430--442, 1997.

\bibitem{jiao2002}
D.~Jiao and J.-M. Jin, ``A general approach for the stability analysis of the
  time-domain finite-element method for electromagnetic simulations,''
  \emph{{IEEE} Trans. Antennas Propag.}, vol.~50, no.~11, pp. 1624--1632, 2002.

\bibitem{jiao2003}
D.~Jiao, J.-M. Jin, E.~Michielssen, and D.~Riley, ``{Time-Domain Finite-Element
  Simulation of Three-Dimensional Scattering and Radiation Problems Using
  Perfectly Matched Layers},'' \emph{{IEEE} Trans. Antennas Propag.}, vol.~51,
  no.~2, pp. 296--305, 2003.

\bibitem{qing2012}
Q.~He, H.~Gan, and D.~Jiao, ``{Explicit Time-Domain Finite-Element Method
  Stabilized for an Arbitrarily Large Time Step},'' \emph{{IEEE} Trans.
  Antennas Propag.}, vol.~60, no.~11, pp. 5240--5250, 2012.

\bibitem{bondeson2012}
A.~Bondeson, T.~Rylander, and P.~Ingelstr{\"o}m, \emph{{Computational
  Electromagnetics}}.\hskip 1em plus 0.5em minus 0.4em\relax Berlin, Germany:
  Springer, 2013.

\bibitem{teixeira2008}
F.~L. Teixeira, ``{Time-Domain Finite-Difference and Finite-Element Methods for
  Maxwell Equations in Complex Media},'' \emph{{IEEE} Trans. Antennas Propag.},
  vol.~56, no.~8, pp. 2150--2166, 2008.

\bibitem{wang2019}
S.~Wang, Y.~Shao, and Z.~Peng, ``{A Parallel-in-Space-and-Time Method for
  Transient Electromagnetic Problems},'' \emph{{IEEE} Trans. Antennas Propag.},
  vol.~67, no.~6, pp. 3961--3973, 2019.

\bibitem{chen2020}
K.~Chen, J.~Liu, M.~Zhuang, Q.~Sun, and Q.~H. Liu, ``{New Mixed SETD and FETD
  Methods to Overcome the Low-Frequency Breakdown Problems by Tree-Cotree
  Splitting},'' \emph{{IEEE} Trans. Microwave Theory Tech.}, vol.~68, no.~8,
  pp. 3219--3228, 2020.

\bibitem{benner2015}
P.~Benner, S.~Gugercin, and K.~Willcox, ``A survey of projection-based model
  reduction methods for parametric dynamical systems,'' \emph{{SIAM} Review},
  vol.~57, no.~4, pp. 483--531, 2015.

\bibitem{hess2013}
M.~W. Hess and P.~Benner, ``{Fast evaluation of time--harmonic Maxwell's
  equations using the reduced basis method},'' \emph{{IEEE} Trans. Microwave
  Theory Tech.}, vol.~61, no.~6, pp. 2265--2274, 2013.

\bibitem{delaRubia2009}
V.~de~la Rubia, U.~Razafison, and Y.~Maday, ``{Reliable Fast Frequency Sweep
  for Microwave Devices via the Reduced-Basis Method},'' \emph{{IEEE} Trans.
  Microw. Theory Tech.}, vol.~57, no.~12, pp. 2923--2937, 2009.

\bibitem{delaRubia2018}
V.~de~la Rubia and M.~Mrozowski, ``{A Compact Basis for Reliable Fast Frequency
  Sweep via the Reduced-Basis Method},'' \emph{{IEEE} Trans. Microw. Theory
  Tech.}, vol.~66, no.~10, pp. 4367--4382, 2018.

\bibitem{feng2019}
L.~Feng and P.~Benner, ``{A New Error Estimator for Reduced-Order Modeling of
  Linear Parametric Systems},'' \emph{{IEEE} Trans. Microwave Theory Tech.},
  vol.~67, no.~12, pp. 4848--4859, 2019.

\bibitem{hochman2014}
A.~Hochman, J.~F. Villena, A.~G. Polimeridis, L.~M. Silveira, J.~K. White, and
  L.~Daniel, ``{Reduced-Order Models for Electromagnetic Scattering
  Problems},'' \emph{{IEEE} Trans. Antennas Propag.}, vol.~62, no.~6, pp.
  3150--3162, 2014.

\bibitem{baltes2017}
R.~Baltes, A.~Schultschik, O.~Farle, and R.~Dyczij-Edlinger, ``{A
  Finite-Element-Based Fast Frequency Sweep Framework Including Excitation by
  Frequency-Dependent Waveguide Mode Patterns},'' \emph{{IEEE} Trans. Microwave
  Theory Tech.}, vol.~65, no.~7, pp. 2249--2260, 2017.

\bibitem{nicolini2019}
J.~L. Nicolini, D.-Y. Na, and F.~L. Teixeira, ``{Model Order Reduction of
  Electromagnetic Particle-in-Cell Kinetic Plasma Simulations via Proper
  Orthogonal Decomposition},'' \emph{{IEEE} Trans. Plasma Sci.}, vol.~47,
  no.~12, pp. 5239--5250, 2019.

\bibitem{codecasa2019}
L.~Codecasa, G.~G. Gentili, and M.~Politi, ``{Exploiting Port Responses for
  Wideband Analysis of Multimode Lossless Devices},'' \emph{{IEEE} Trans.
  Microw. Theory Tech.}, vol.~68, no.~2, pp. 555--563, 2020.

\bibitem{jiao2020}
L.~Xue and D.~Jiao, ``{Rapid Modeling and Simulation of Integrated Circuit
  Layout in Both Frequency and Time Domains From the Perspective of Inverse},''
  \emph{{IEEE} Trans. Microwave Theory Tech.}, vol.~68, no.~4, pp. 1270--1283,
  2020.

\bibitem{chellappa2021}
S.~Chellappa, L.~Feng, V.~de~la Rubia, and P.~Benner, \emph{{Adaptive
  Interpolatory {MOR} by Learning the Error Estimator in the Parameter
  Domain}}.\hskip 1em plus 0.5em minus 0.4em\relax Cham: Springer International
  Publishing, 2021, pp. 97--117.

\bibitem{ziegler2023}
A.~Ziegler, N.~Georg, W.~Ackermann, and S.~Schöps, ``{Mode Recognition by
  Shape Morphing for Maxwell's Eigenvalue Problem in Cavities},'' \emph{{IEEE}
  Trans. Antennas Propag.}, vol.~71, no.~5, pp. 4315--4325, 2023.

\bibitem{kappesser2024}
M.~Kappesser, A.~Ziegler, and S.~Schöps, ``{Reduced Basis Approximation for
  Maxwell's Eigenvalue Problem and Parameter-Dependent Domains},'' \emph{IEEE
  Trans. Magn.}, vol.~60, no.~3, 2024.

\bibitem{szypulski2020}
D.~{Szypulski}, G.~{Fotyga}, V.~{de la Rubia}, and M.~{Mrozowski}, ``{A
  Subspace-Splitting Moment-Matching Model-Order Reduction Technique for Fast
  Wideband FEM Simulations of Microwave Structures},'' \emph{{IEEE} Trans.
  Microwave Theory Tech.}, vol.~68, no.~8, pp. 3229--3241, 2020.

\bibitem{rewienski2016}
M.~Rewienski, A.~Lamecki, and M.~Mrozowski, ``{Greedy Multipoint Model-Order
  Reduction Technique for Fast Computation of Scattering Parameters of
  Electromagnetic Systems},'' \emph{{IEEE} Trans. Microwave Theory Tech.},
  vol.~64, no.~6, pp. 1681--1693, 2016.

\bibitem{fotyga2018}
G.~Fotyga, M.~Czarniewska, A.~Lamecki, and M.~Mrozowski, ``{Reliable Greedy
  Multipoint Model-Order Reduction Techniques for Finite-Element Analysis},''
  \emph{{IEEE} Antennas Wirel. Propag. Lett.}, vol.~17, no.~5, pp. 821--824,
  2018.

\bibitem{li2018}
K.~Li, T.-Z. Huang, L.~Li, and S.~Lanteri, ``{A reduced-order DG formulation
  based on POD method for the time-domain Maxwell’s equations in dispersive
  media},'' \emph{J. Comput. Appl. Math.}, vol. 336, pp. 249--266, 2018.

\bibitem{li2018ieee}
K.~Li, T.-Z. Huang, L.~Li, S.~Lanteri, L.~Xu, and B.~Li, ``{A Reduced-Order
  Discontinuous Galerkin Method Based on POD for Electromagnetic Simulation},''
  \emph{{IEEE} Trans. Antennas Propag.}, vol.~66, no.~1, pp. 242--254, 2018.

\bibitem{li2019}
K.~Li, T.-Z. Huang, L.~Li, and S.~Lanteri, ``{POD-based model order reduction
  with an adaptive snapshot selection for a discontinuous Galerkin
  approximation of the time-domain Maxwell's equations},'' \emph{J. Comput.
  Phys.}, vol. 396, pp. 106--128, 2019.

\bibitem{li2022}
K.~Li, T.-Z. Huang, L.~Li, Y.~Zhao, and S.~Lanteri, ``{A non-intrusive model
  order reduction approach for parameterized time-domain Maxwell's
  equations},'' \emph{{Discrete Contin. Dyn. Syst. - B}}, vol.~28, no.~1, pp.
  449--473, 2022.

\bibitem{li2023}
K.~Li, Y.~Li, L.~Li, and S.~Lanteri, ``Surrogate modeling of time-domain
  electromagnetic wave propagation via dynamic mode decomposition and radial
  basis function,'' \emph{J. Comput. Phys.}, vol. 491, p. 112354, 2023.

\bibitem{li2023nmpde}
K.~Li, T.-Z. Huang, L.~Li, and S.~Lanteri, ``{Simulation of the interaction of
  light with 3-D metallic nanostructures using a proper orthogonal
  decomposition-Galerkin reduced-order discontinuous Galerkin time-domain
  method},'' \emph{Numer. Methods Partial Differ. Equ.}, vol.~39, no.~2, pp.
  932--954, 2023.

\bibitem{luo2016}
Z.~Luo and J.~Gao, ``{A POD reduced-order finite difference time-domain
  extrapolating scheme for the 2D Maxwell equations in a lossy medium},''
  \emph{J. Math. Anal. Appl.}, vol. 444, no.~1, pp. 433--451, 2016.

\bibitem{nayak2024}
I.~Nayak, F.~L. Teixeira, and R.~J. Burkholder, ``{On-the-Fly Dynamic Mode
  Decomposition for Rapid Time-Extrapolation and Analysis of Cavity
  Resonances},'' \emph{{IEEE} Trans. Antennas Propag.}, vol.~72, no.~1, pp.
  131--146, 2024.

\bibitem{nicolini2023}
J.~L. Nicolini, F.~L. Teixeira, and R.~J. Burkholder, ``Reduced-order mode
  discovery in arbitrary cavities via proper orthogonal decomposition,'' in
  \emph{2023 IEEE International Symposium on Antennas and Propagation and
  USNC-URSI Radio Science Meeting (USNC-URSI)}, 2023, pp. 837--838.

\bibitem{yu2024}
M.~Yu and K.~Li, ``{A data-driven reduced-order modeling approach for
  parameterized time-domain Maxwell's equations},'' \emph{Netw. Heterog.
  Media}, vol.~19, no.~3, pp. 1309--1335, 2024.

\bibitem{leclainche2017}
S.~Le~Clainche and J.~M. Vega, ``{Higher Order Dynamic Mode Decomposition},''
  \emph{SIAM J. Appl. Dyn. Syst.}, vol.~16, no.~2, pp. 882--925, 2017.

\bibitem{nayak2024plasma}
I.~Nayak, F.~L. Teixeira, D.-Y. Na, M.~Kumar, and Y.~A. Omelchenko,
  ``Accelerating particle-in-cell kinetic plasma simulations via reduced-order
  modeling of space-charge dynamics using dynamic mode decomposition,''
  \emph{Phys. Rev. E}, vol. 109, p. 065307, 2024.

\bibitem{medeiros2024TAP}
\BIBentryALTinterwordspacing
R.~Medeiros and V.~de~la Rubia, ``{A Reduced Order Model for Finite Element
  Method in Time Domain Electromagnetic Simulations},'' 2024. [Online].
  Available: \url{https://arxiv.org/abs/2410.16572}
\BIBentrySTDinterwordspacing

\bibitem{jane2023}
E.~Jan{\'e}, R.~Medeiros, F.~Varas, and M.~Higuera, ``{A Time-Adaptive Order
  Reduction Technique for the Doyle-Fuller-Newman Electrochemical Model of
  Lithium-Ion Batteries},'' \emph{J. Electrochem. Soc.}, vol. 170, no.~3, p.
  030539, 2023.

\bibitem{doyle1993}
M.~Doyle, T.~F. Fuller, and J.~Newman, ``{Modeling of Galvanostatic Charge and
  Discharge of the Lithium/Polymer/Insertion Cell},'' \emph{J. Electrochem.
  Soc.}, vol. 140, no.~6, pp. 1526--1533, 1993.

\bibitem{medeiros2024}
R.~Medeiros, E.~Jan{\'e}, F.~Varas, and M.~Higuera, ``Battery cell optimisation
  using time- and parameter-adaptive reduced order models,'' \emph{Comput.
  Math. Appl.}, vol. 161, pp. 137--154, 2024.

\bibitem{willcox2002}
K.~Willcox and J.~Peraire, ``{Balanced Model Reduction via the Proper
  Orthogonal Decomposition},'' \emph{AIAA Journal}, vol.~40, no.~11, pp.
  2323--2330, 2002.

\bibitem{kirsch2014}
A.~Kirsch and F.~Hettlich, \emph{{The Mathematical Theory of Time-Harmonic
  Maxwell's Equations: Expansion-, Integral-, and Variational Methods}}.\hskip
  1em plus 0.5em minus 0.4em\relax Cham: Springer, 2014.

\bibitem{monk2003}
P.~Monk, \emph{{Finite Element Methods for Maxwell's Equations}}.\hskip 1em
  plus 0.5em minus 0.4em\relax Oxford: Oxford University Press, 2003.

\bibitem{jin2014}
J.-M. Jin, \emph{{The Finite Element Method in Electromagnetics}},
  3rd~ed.\hskip 1em plus 0.5em minus 0.4em\relax Hoboken, NJ, USA: John Wiley
  \& Sons, 2014.

\bibitem{jiao2012}
Q.~He, H.~Gan, and D.~Jiao, ``{Explicit Time-Domain Finite-Element Method
  Stabilized for an Arbitrarily Large Time Step},'' \emph{{IEEE} Trans.
  Antennas Propag.}, vol.~60, no.~11, pp. 5240--5250, 2012.

\bibitem{van2004}
T.~Van and A.~Wood, ``{A Time-Domain Finite Element Method for Maxwell's
  Equations},'' \emph{SIAM J. Numer. Anal.}, vol.~42, no.~4, pp. 1592--1609,
  2004.

\bibitem{sirovich1987}
L.~Sirovich, ``{Turbulence and the dynamics of coherent structures. I. Coherent
  structures},'' \emph{Quart. Appl. Math.}, vol.~45, no.~3, pp. 561--571, 1987.

\bibitem{terragni2011}
F.~Terragni, E.~Valero, and J.~M. Vega, ``{Local POD plus Galerkin projection
  in the unsteady lid-driven cavity problem},'' \emph{SIAM J. Sci. Comput.},
  vol.~33, no.~6, pp. 3538--3561, 2011.

\bibitem{rapun2015}
M.-L. Rap{\'u}n, F.~Terragni, and J.~M. Vega, ``{Adaptive POD-based
  low-dimensional modeling supported by residual estimates},'' \emph{Int. J.
  Numer. Methods Eng.}, vol. 104, no.~9, pp. 844--868, 2015.

\bibitem{grivet2015}
S.~Grivet-Talocia and B.~Gustavsen, \emph{{Passive Macromodeling: Theory and
  Applications}}.\hskip 1em plus 0.5em minus 0.4em\relax Hoboken, NJ: John
  Wiley \& Sons, 2014.

\bibitem{nedelec1980}
J.-C. N{\'e}d{\'e}lec, ``Mixed finite elements in $\mathbb{R}^3$,''
  \emph{Numer. Math.}, vol.~35, no.~3, pp. 315--341, 1980.

\bibitem{ingelstrom2006}
P.~Ingelstrom, ``A new set of h(curl)-conforming hierarchical basis functions
  for tetrahedral meshes,'' \emph{{IEEE} Trans. Microw. Theory Tech.}, vol.~54,
  no.~1, pp. 106--114, 2006.

\bibitem{geuzaine2009}
C.~Geuzaine and J.-F. Remacle, ``Gmsh: A {3-D} finite element mesh generator
  with built-in pre-and post-processing facilities,'' \emph{Int. J. Numer.
  Methods Eng.}, vol.~79, no.~11, pp. 1309--1331, 2009.

\bibitem{lou2005}
Z.~Lou and J.-M. Jin, ``Modeling and simulation of broad-band antennas using
  the time-domain finite element method,'' \emph{{IEEE} Trans. Antennas
  Propag.}, vol.~53, no.~12, pp. 4099--4110, 2005.

\bibitem{li2005}
B.~Li and K.~W. Leung, ``Strip-fed rectangular dielectric resonator antennas
  with/without a parasitic patch,'' \emph{{IEEE} Trans. Antennas Propag.},
  vol.~53, no.~7, pp. 2200--2207, 2005.

\bibitem{parellada2024}
I.~Parellada-Serrano, M.~Pérez-Escribano, C.~Molero, P.~Padilla, and V.~de~la
  Rubia, ``{Three-Dimensional Fully Metallic Dual-Polarization
  Frequency-Selective Surface Design Using Coupled-Resonator Circuit
  Information},'' \emph{{IEEE} Trans. Antennas Propag.}, vol.~72, no.~3, pp.
  2932--2937, 2024.

\bibitem{parellada2024data}
------, ``{AutoCAD Files Including Main Dimensions},'' 2023.

\bibitem{mouris2020}
B.~A. Mouris, A.~Fernández-Prieto, R.~Thobaben, J.~Martel, F.~Mesa, and
  O.~Quevedo-Teruel, ``{On the Increment of the Bandwidth of Mushroom-Type EBG
  Structures With Glide Symmetry},'' \emph{{IEEE} Trans. Microw. Theory Tech.},
  vol.~68, no.~4, pp. 1365--1375, 2020.

\bibitem{koivurova2023}
M.~Koivurova, C.~W. Robson, and M.~Ornigotti, ``Time-varying media, relativity,
  and the arrow of time,'' \emph{Optica}, vol.~10, no.~10, pp. 1398--1406,
  2023.

\bibitem{pacheco-pena2024}
V.~Pacheco-Peña and N.~Engheta, ``{Spatiotemporal cascading of dielectric
  waveguides [Invited]},'' \emph{Opt. Mater. Express}, vol.~14, no.~4, pp.
  1062--1073, 2024.

\bibitem{brunton2019}
S.~L. Brunton and J.~N. Kutz, \emph{{Data-driven science and engineering:
  Machine learning, dynamical systems, and control}}.\hskip 1em plus 0.5em
  minus 0.4em\relax Cambridge: Cambridge University Press, 2019.

\bibitem{lucia2004}
D.~J. Lucia, P.~S. Beran, and W.~A. Silva, ``Reduced-order modeling: new
  approaches for computational physics,'' \emph{Prog. Aerosp. Sci.}, vol.~40,
  no.~1, pp. 51--117, 2004.

\bibitem{chaturantabut2012}
S.~Chaturantabut and D.~C. Sorensen, ``{A State Space Error Estimate for
  POD-DEIM Nonlinear Model Reduction},'' \emph{SIAM J. Numer. Anal.}, vol.~50,
  no.~1, pp. 46--63, 2012.

\bibitem{hesthaven2016}
J.~S. Hesthaven, G.~Rozza, B.~Stamm \emph{et~al.}, \emph{{Certified Reduced
  Basis Methods for Parametrized Partial Differential Equations}}.\hskip 1em
  plus 0.5em minus 0.4em\relax Springer, 2016.

\bibitem{feng2017}
L.~Feng, M.~Mangold, and P.~Benner, ``{Adaptive POD-DEIM basis construction and
  its application to a nonlinear population balance system},'' \emph{AIChE
  Journal}, vol.~63, no.~9, pp. 3832--3844, 2017.

\end{thebibliography}

\end{document}